%% file: main.tex
\documentclass[10pt,journal,compsoc]{IEEEtran}
\ifCLASSOPTIONcompsoc
  \usepackage[nocompress]{cite}
\else
  \usepackage{cite}
\fi
\ifCLASSINFOpdf
\else
\fi

\usepackage{booktabs}
\usepackage{xcolor}
\usepackage{multirow}
\usepackage{graphicx,enumitem}
\usepackage[numbers]{natbib}
\usepackage{algorithm}
\usepackage{algorithmic}
\usepackage{amsmath}
\usepackage{amssymb}
\usepackage{dsfont}
\usepackage{array}
\usepackage{booktabs}
\usepackage{threeparttable}
\usepackage{subfigure}

\newcommand{\tabincell}[2]{\begin{tabular}{@{}#1@{}}#2\end{tabular}}
\newtheorem{definition}{Definition}
\begin{document}
%
\title{AIM: Automatic Interaction Machine for Click-Through Rate Prediction}

\author{Chenxu~Zhu, Bo~Chen, Weinan~Zhang, Jincai~Lai, Ruiming~Tang, Xiuqiang~He, Zhenguo~Li and Yong~Yu 
	\IEEEcompsocitemizethanks{\IEEEcompsocthanksitem This paper is an extended version of AutoFIS ~\cite{autofis} which has been accepted for the presentation at 26th ACM SIGKDD International Conference on Knowledge Discovery \& Data Mining.
	\IEEEcompsocthanksitem Chenxu Zhu, Weinan Zhang and Yong Yu are with the Shanghai Jiao Tong University. E-mail: \{zhuchenxv, wnzhang, yyu\}@sjtu.edu.cn.
	\IEEEcompsocthanksitem Bo Chen, Jincai Lai, Ruiming Tang and Xiuqiang He are with the Huawei Noah's Ark Lab. E-mail: \{chenbo116, laijincai, tangruiming, hexiuqiang1, Li.Zhenguo\}@huawei.com.
	\IEEEcompsocthanksitem Corresponding author: Weinan Zhang.}}

%
%

\markboth{IEEE TRANSACTIONS ON KNOWLEDGE AND DATA ENGINEERING}%
{Shell \MakeLowercase{\textit{et al.}}: Bare Demo of IEEEtran.cls for Computer Society Journals}
%




\newcommand{\chenxu}[1]{{\bf \color{red} [[Chenxu says ``#1'']]}}
\newcommand{\chenbo}[1]{{\bf \color{green} [[Chenbo says ``#1'']]}}
\newcommand{\ruiming}[1]{{\bf \color{orange} [[Ruiming says ``#1'']]}}
\newcommand{\weinan}[1]{{\bf \color{blue} [[Weinan says ``#1'']]}}

\input{abstract.tex}
\maketitle

\IEEEdisplaynontitleabstractindextext

%
\IEEEpeerreviewmaketitle

\begin{figure*}
	\centering
	\vspace{-1.0em}
	\includegraphics[width=0.8 \textwidth]{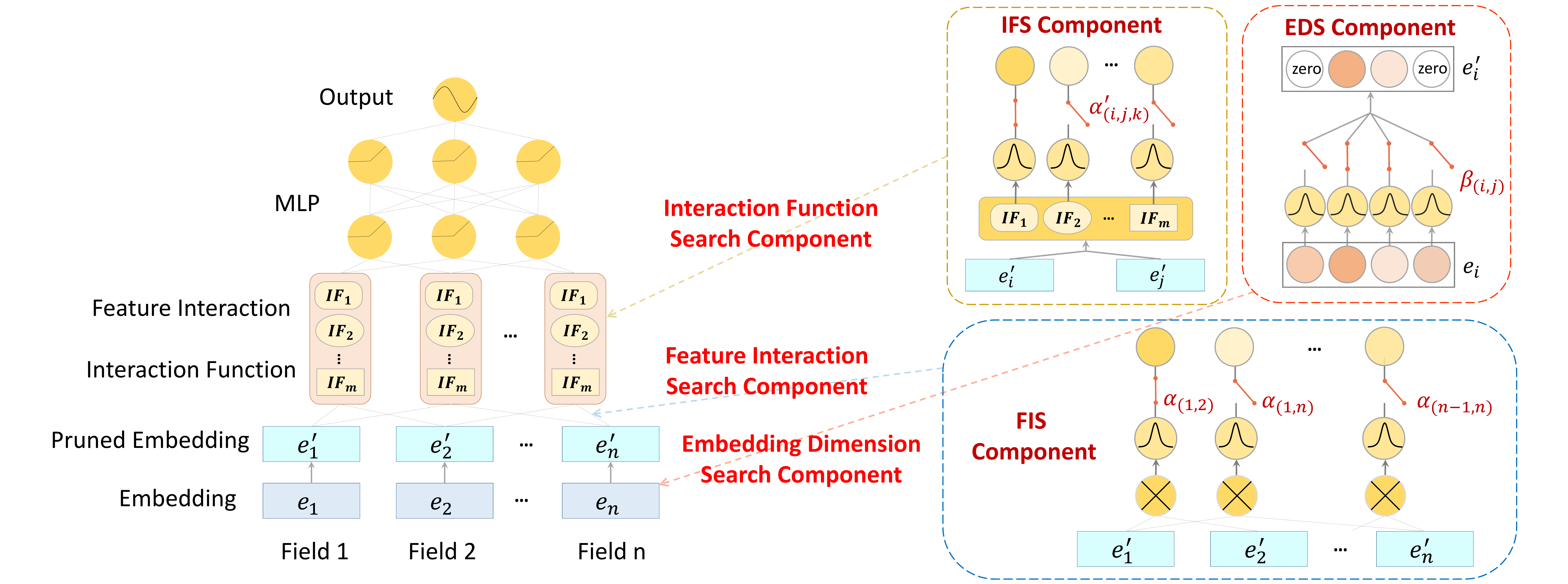}
	\vspace{-1.1em}
	\caption{Architectures of Automatic Interaction Machine (AIM).}
	\vspace{-1.4em}
	\label{fig:aim}
\end{figure*}
\input{intro.tex}
\input{related_work}

\input{model.tex}

\input{exp.tex}

\input{conclusion.tex}

\section*{acknowledgements}
The SJTU team is supported by ``New Generation of AI 2030'' Major Project (2018AAA0100900), Shanghai Municipal Science and Technology Major Project (2021SHZDZX0102) and National Natural Science Foundation of China (62076161, 62177033). The work is also sponsored by Huawei Innovation Research Program.

\ifCLASSOPTIONcaptionsoff
  \newpage
\fi



\bibliographystyle{IEEEtranN}
\bibliography{IEEEabrv,autofifs}

\end{document}

%% file: abstract.tex
\IEEEtitleabstractindextext{%
\begin{abstract}

Feature embedding learning and feature interaction modeling are two crucial components of deep models for Click-Through Rate (CTR) prediction in recommender systems. Most existing deep CTR models suffer from the following three problems. First, feature interactions are either manually designed or simply enumerated. However, not all the feature interactions are useful for the prediction task and useless feature interactions may introduce noisy signals 
thus causing overfitting. 
Second, all the feature interactions are modeled with an identical interaction function, 
whereas different interaction functions introduce different inductive biases to better capture various feature interaction patterns.
Third, in most existing models, different features share the same embedding size. However, model size can be further optimized without sacrificing performance by differentiating embedding sizes for individual features, as the amount of information contained in each feature varies much. 
To address the three issues mentioned above, we propose \textbf{Automatic Interaction Machine (AIM)} with three core components, namely, Feature Interaction Search (FIS), Interaction Function Search (IFS) and Embedding Dimension Search (EDS), respectively.
To tackle the first problem, FIS component  automatically identifies different orders of essential feature interactions with useless ones pruned.
Taking care of the second problem, IFS component selects appropriate interaction functions for each individual feature interaction in a learnable way. Moreover, to avoid learning conflict among different interaction functions, IFS proposes function-wise embeddings via performing multiple embeddings for each feature, where each feature embedding corresponds to one possible interaction function.
However, utilizing multiple embeddings for each feature may make the model size affordably large if we keep the same embedding size as utilizing shared embedding (i.e., each feature shares the same embedding for different interaction functions). 
To solve this third problem, EDS automatically selects proper embedding size for each feature. Such a flexible embedding size adaptation is able to reduce the large amount of embedding parameters introduced by function-wise embeddings.
Offline experiments on three large-scale datasets (two public benchmarks, one private dataset) validate that AIM can significantly improve various FM-based models. AIM has been deployed in the recommendation service of a mainstream app market, where a three-week online A/B test demonstrated the superiority of AIM, improving DeepFM model by 4.4\% in terms of CTR. 


\end{abstract}

\begin{IEEEkeywords}
CTR Prediction, Feature Interaction, Neural Architecture Search, Factorization Machine, Recommender Systems
\end{IEEEkeywords}}

%% file: intro.tex
\section{Introduction}\label{sec:intro}



Click-Through Rate (CTR) prediction, which aims to predict the probability of the user clicking on the recommended items (e.g., music, advertisement), plays a core role in recommender systems. The estimated CTR may influence the subsequent recommendation decision-makings such as item ranking and ad placement. Accurate CTR prediction not only improves the user experience but also boosts the profit for the service providers. Since 2016, deep learning has been introduced to CTR prediction due to its high capacity of modeling high-order patterns and end-to-end learning manner \cite{zhang2021deep}. So far, various deep CTR models have been deployed in the recommendation services of industrial companies, such as Wide \& Deep~\cite{widedeep} in Google Play, DeepFM~\cite{deepfm} in Huawei AppGallery and DIN~\cite{din} in Taobao.


Most of the existing deep CTR prediction models consist of two key components: feature embedding learning and feature interaction modeling. Feature embeddings are learned via mapping categorical features into low-dimensional embedding vectors (short for embeddings). 
Feature interactions are learned by utilizing some functions to model the relationship among a set of feature embeddings. Many research works focus on designing interaction functions (or more generally, network architectures) to better capture feature interactions.

At the early stage, Deep Neural Network models~\cite{dnnyoutube,fnn,widedeep} are proposed to model feature interactions implicitly with the multi-layer perceptron (MLP) built on the feature embeddings. In theory, a DNN could explore arbitrary feature interactions according to its universal approximation property~\cite{DNN_universial}. However, there is no guarantee that a DNN naturally converges to any expected function using gradient-based optimization. Recent works prove the insensitive gradient issue of DNN when the target is a large collection of uncorrelated functions~\cite{pin,fail_dnn}. Simple DNN models may not find the proper feature interactions, including the simple inner product \cite{rendle2020neural}. 
Therefore, various complicated architectures have been proposed, such as DIN~\cite{din}, DeepFM~\cite{deepfm} and PNN~\cite{pnn}.
\textbf{Factorization Models} (specified in Definition 1), such as FM~\cite{fm}, DeepFM, PNN, AFM~\cite{afm}, NFM~\cite{nfm}, have been proposed to adopt an embedding layer and an inner-product feature extractor to respectively learn feature embedding and feature interactions.


However, there are three significant downsides in most of these existing models.
First, all these models simply enumerate all feature interactions or require human efforts to identify important feature interactions. 
The former always brings large memory and computation cost to the model and is difficult to extend into high-order interactions. Besides, useless interactions may bring unnecessary noise to cause overfitting and complicate the training process~\cite{fail_dnn}. The latter, such as identifying important interactions manually in Wide \& Deep~\cite{widedeep}, is of high labor cost and risks missing some counter-intuitive (but important) interactions.

Second, all the feature interactions with different complexity are modeled with the same specific interaction function (IF), such as inner product~\cite{pnn,deepfm,afm} or a predefined sub-network~\cite{pin}. 
Nevertheless, using different interaction functions is conducive to better modeling various feature interactions under different mapping sub-spaces, especially for higher-order feature interactions as pointed out in AutoFeature~\cite{autofeature}.

Third, different features share the same embedding size to represent their embeddings. As pointed out in some remarkable works \cite{auto-embed-autodim,auto-embed-autoemb,auto-embed-ESAPN,auto-embed-nis}, such 
a strategy may lead to memory inefficiency. More specifically, allocating the same embedding size to all features may lose the information of high predictive features while waste memory on non-predictive features. Therefore, model size can be further optimized without sacrificing performance by differentiating embedding sizes for individual features, i.e., assigning large embedding size to high predictive features, while assigning small embedding size to low predictive ones.


Some recent works improve the existing models by solving part of the above three problems. AutoFIS~\cite{autofis} and AutoGroup~\cite{autogroup} identify important feature interactions while performing identical IF for all the selected feature interactions with uniform embedding size (namely, solving the first problem only). AutoFeature~\cite{autofeature} solves the first and second problems by automatically searching proper network architectures to model different feature interactions via an evolutionary algorithm with the Naive Bayes tree. However, high complexity hinders the application of this method to large-scale industrial scenarios.
To solve the third problem, a bunch of previous works \cite{auto-embed-autodim,auto-embed-autoemb,auto-embed-ESAPN,auto-embed-nis} utilize various techniques to find proper embedding sizes for each feature automatically, but the first two issues about feature interactions are totally ignored in these works. In other words, none of these advanced work solves all these three important problems in a unified framework.

To fill this gap, in this paper, we propose a unified framework called \textbf{Automatic Interaction Machine (AIM}) with three core components, i.e., \textit{Feature Interaction Search (FIS)}, \textit{Interaction Function Search (IFS)} and \textit{Embedding Dimension Search (EDS)}, which are elaborated in Figure~\ref{fig:aim}.

First, FIS component automatically learns which feature interactions are essential. Specifically, we introduce a \textit{gate} (in open or closed status) for each feature interaction to control whether its output should be passed to the next layer. 
In previous works, the status of the gates are either specified beforehand by expert knowledge~\cite{widedeep} or set as all open~\cite{deepfm,xdeepfm}. From a data-driven point of view, whether to open or close a gate should depend on the contribution to the final prediction. Apparently, those contributing little should be closed to prevent the learning procedure from introducing extra noise. However, it is an NP-Hard problem to find the optimal set of open gates for model performance, as we face an incredibly huge space to search ($2^{\mathcal{C}_{n}^{2}}$ with $n$ fields, even if we only consider $2^{nd}$-order feature interactions). 
To make the search efficient in such a huge space, FIS component relaxes the choices of gates to be continuous such that gradient-based optimizations can be performed.
Furthermore, FIS component is also able to select high-order feature interactions. To make the search process efficient, we exploit the selected low-order interaction to first restrict high-order interactions into a small candidate pool heuristically and then search from this pool rigorously, achieving $O(n^2)$ complexity.



As the second component of AIM, IFS can select appropriate IF for each essential feature interaction. 
To achieve this goal, a \textit{gate} to each interaction-IF pair is needed.
It can be observed that the search space of IFS is larger than that of FIS, which reaches to $2^{m\mathcal{C}_{n}^{2}}$ (with $n$ fields and $m$ IFs), if we only consider $2^{nd}$-order feature interactions. To avoid learning conflict among different IFs, IFS proposes to utilize function-wise embeddings, i.e., multiple embeddings learned for each feature, where each of such embeddings corresponds to one possible IF\footnote{\scriptsize{As the opposite, we refer \textit{shared embedding} as the case that each feature shares the same embedding for different IFs.}}. 


As function-wise embeddings are used in IFS, the model size grows significantly (about $m$ times than shared embedding). 
The third component, EDS, is proposed to optimize the model size without sacrificing performance by assigning small embedding sizes to low predictive features, which is achieved by pruning redundant embedding dimensions for each feature. Similar to FIS and IFS components, \textit{gates} are also introduced, where a \textit{gate} is annotated to each dimension of a feature embedding. In the end, the dimensions with open gates are recognized as important dimensions while the ones with closed gates are pruned.


Inspired by the recent work DARTS~\cite{liu2018darts, MiLeNAS,li2019stacnas} for neural architecture search, instead of searching over a discrete set of candidate gates (i.e., feature interactions in FIS, interaction-IF pairs in IFS and embedding dimensions in EDS), we relax the choices to be continuous by introducing a set of architecture parameters (one for each gate) so that the relative importance can be learned by gradient descent. 
The architecture parameters are jointly optimized with neural network weights by GRDA optimizer~\cite{chao2019generalization}, an optimizer that is easy to produce a sparse solution, so that the training process can automatically abandon unimportant feature interactions, useless interaction-IF pairs and unnecessary embedding dimensions with zero values as the architecture parameters and keep those important ones. Finally, we re-train the model with the selected feature interactions, interaction-IF pairs as well as embedding dimensions.

Extensive experiments are conducted on three large-scale datasets and the experimental results demonstrate that AIM can significantly improve the CTR prediction performance in terms of AUC and log loss. 
Besides, as AIM can remove about 80\% of $2^{nd}$-order interactions, our model can consistently achieve improvement on inference efficiency. AIM is able to model high-order feature interactions in a novel way with quadratic complexity. Experimental results show that with about 1\%--5\% of $3^{rd}$-order feature interactions and about 0.03\%--0.6\% of $4^{th}$-order feature interactions selected, the AUC of factorization models can be improved by 0.1\%--0.4\% without introducing much inference cost. 
Furthermore, AIM has been deployed in the recommendation service of a mainstream app market. From a three-week online A/B test, AIM achieves $4.4\%$ CTR improvement over the production model DeepFM, which contributes a significant business revenue growth. To summarize, the main contributions of this paper can be highlighted as follows:
\begin{itemize}[leftmargin = 12 pt]
    \item We propose \textbf{AIM} with three core components (namely, FIS, IFS and EDS components) to select significant feature interactions, appropriate IFs and necessary embedding dimensions automatically in a unified framework. 
	\item \textit{Gates} are introduced in FIS, IFS and EDS components, with open (closed) status representing important (unimportant) candidates. We relax discrete selection of open gates to be continuous by introducing a set of architecture parameters that can be jointly optimized with neural network weights by GRDA optimizer.
	
	\item Offline experiments on three large-scale datasets demonstrate the superior performance of AIM.  
	A three-week online A/B test in the recommendation service of a mainstream app market shows that AIM improves DeepFM model by 4.4\% on average in terms of CTR.
\end{itemize}

%% file: related_work.tex
\section{Related Work}


\subsection{CTR Models}\label{sec:related-work-ctr-models}

One core of CTR models is to extract effective low-order and high-order feature interactions, which is also one of the optimization targets of our work. Therefore, in this section, we discuss the CTR models that focus on learning feature interactions effectively.

FM~\cite{fm} projects each feature into a low-dimensional vector and models pair-wise feature interactions by inner product, which works well for sparse data. FFM~\cite{ffm} extends FM, by further assigning each feature with multiple vectors to interact with features from other fields. Despite the significant improvement over FM, FFM introduces much more parameters and suffers from overfitting issues. HOFM~\cite{HOFM} introduces the ANOVA kernel to approximate high-order feature interactions. However, it is shown in~\cite{afm} that HOFM achieves only marginal improvement over FM whereas using much more parameters. 

Recently, deep learning models have achieved state-of-the-art performance on some public CTR prediction benchmarks~\cite{fgcnn,crossnet}. As a powerful approach to learning feature representation, deep learning models have the potential to learn sophisticated feature interactions. 
Wide \& Deep~\cite{widedeep} jointly trains a wide model for artificial features and a deep model for raw features. Several models use MLP to improve FM, such as AFM~\cite{afm}, NFM~\cite{nfm} and DeepFM~\cite{deepfm}. DeepFM uses an FM layer to replace the wide component in Wide \& Deep. PNN~\cite{pnn} uses MLP to model high-order implicit information over the feature embeddings and interaction of FM layer while 
PIN~\cite{pin} introduces a network-in-network architecture to model pairwise feature interactions with sub-networks rather than simple inner product operations in PNN and DeepFM. Due to the curse of dimensionality, DeepFM, PNN and PIN cannot explicitly model high-order feature interactions, which limits the further improvement of the model performance. xDeepFM~\cite{xdeepfm} uses a CIN structure to enumerate and compress all feature interactions for modeling explicit interactions. However, it uses so many parameters that great challenges are posed to identify important feature interactions in the huge combination space.

As stated earlier, all the above mentioned CTR models suffer from three limitations: (1) enumerating all the feature interactions or requiring human efforts; (2) utilizing the same IF to model all the feature interactions; (3) assigning the same embedding size to all the features. Our proposed AIM improves these models by solving such three downsides with AutoML techniques.



\subsection{AutoML for CTR Models}  

AutoML for CTR models has been an active research area and there exist some works using AutoML techniques to devise automated methods for architecture design including embedding dimension search and feature interaction search. 

Existing works leverage the AutoML to optimize the embeddings by searching embedding sizes for different features adaptively and automatically. NIS~\cite{auto-embed-nis} and ESAPN~\cite{auto-embed-ESAPN} perform a hard selection strategy and use reinforcement learning (RL) to search for mixed feature embedding sizes automatically. On the contrary, soft selection strategies based on differentiable search algorithms (e.g., DARTS~\cite{liu2018darts}) are proposed in AutoEmb~\cite{auto-embed-autoemb} and AutoDim~\cite{auto-embed-autodim} by summing over the embeddings of the candidate sizes with learnable weights. The former leverages a controller network with Softmax layer to search the weights of different embedding sizes while the latter uses the Gumbel-softmax operation~\cite{gumbel}.

As stated earlier, for CTR models, feature interaction modeling is an important component, where the main tasks are to find which feature interactions are useful and decide how the interaction should be modeled. Some research works are proposed to explore these two questions.
AutoFIS~\cite{autofis} automatically identifies and then selects important feature interactions for factorization models with a set of learnable architecture parameters. 
Besides, AutoGroup~\cite{autogroup} proposes
to generate some groups of features, such that their interactions of a given order are effective, which enables high-order feature interaction modeling. 
To determine suitable IFs, SIF~\cite{autoML_for_CF} exploits the DARTS method to search proper interaction functions for matrix factorization. AutoFeature~\cite{autofeature} applies an evolutionary algorithm with the Naive Bayes tree that recursively reduces the search space of IFs for feature interactions. However, AutoFeature trains excessive models for different structures to select the most appropriate structure, which requires a long period to search and a much high computational resource. 

However, the above mentioned works solve one or two limitations (the three limitations are mentioned in Section~\ref{sec:intro} and also in Section~\ref{sec:related-work-ctr-models}) of the existing CTR models. None of them is able to solve all the three important issues in a unified framework. Our proposed AIM designs a gate-based unified framework and formulates the problem of searching proper embedding size, effective feature interactions as well as appropriate IFs as a continuous searching problem by incorporating architecture parameters that can be jointly optimized with neural network weights.

%% file: model.tex
\begin{figure*}
	\centering
	\vspace{-0.5em}
	\includegraphics[width=0.85 \textwidth]{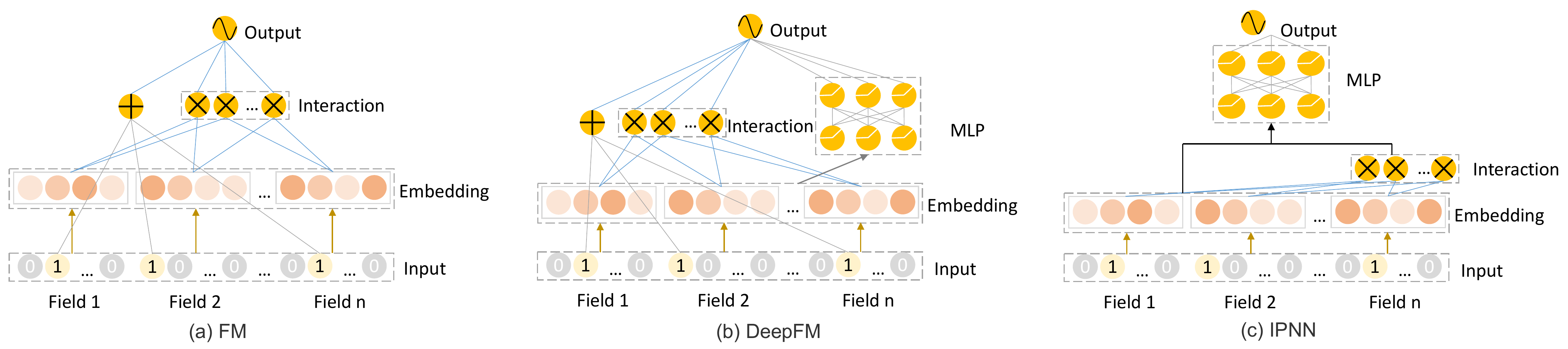}
	\vspace{-0.9em}
	\caption{ Architectures of FM, DeepFM and IPNN.}
	\vspace{-1.4em}
	\label{fig:fm_deepfm_ipnn}
\end{figure*}

\section{METHODOLOGY}
In this section, we first formally define a family of popular CTR models (Factorization Model), with which our proposed AIM is able to work collaboratively to improve their performance. Then Automatic Interaction Machine (AIM) with three core components, namely, Feature Interaction Search (FIS), Interaction Function Search (IFS) and Embedding Dimension Search (EDS) is proposed to select significant feature interactions, appropriate IFs and necessary embedding dimensions automatically in a unified framework, as shown in Figure~\ref{fig:aim}.  Finally we discuss some concrete training details.


\subsection{Factorization Model (Base Model)}
First, we define factorization models:
\begin{definition}
\textbf{Factorization models} are the models where the interaction of several embeddings from different features is modeled into a real number by some functions such as inner product or neural network.
\end{definition}

FM~\cite{fm}, DeepFM~\cite{deepfm} and IPNN~\cite{pnn} (IPNN is one kind of PNN when the IF is inner product)
are popular CTR models, which are in the family of factorization model. Therefore, we take FM, DeepFM, and IPNN as instances to formulate our algorithm and explore the performance on various datasets. Figure~\ref{fig:fm_deepfm_ipnn} presents the architectures of FM, DeepFM and IPNN models.
FM consists of a \emph{feature embedding layer} and a \emph{feature interaction layer}. Besides these two layers, DeepFM and IPNN include an extra layer: \emph{MLP layer}. The difference between DeepFM and IPNN is that feature interaction layer and MLP layer work in parallel in DeepFM while ordered in sequence in IPNN.

In the subsequent subsections, we will first elaborate on these layers. Then the details of how our proposed AIM working on the feature embedding layer and feature interaction layer are elaborated, i.e., selecting crucial feature interaction with proper interaction functions (IFs) and appropriate embedding sizes based on architecture parameters. 


\textbf{Feature Embedding Layer.}
In most CTR prediction tasks, data is collected in a multi-field categorical form\footnote{Features in the numerical form are usually transformed into categorical form by bucketing.}. A typical data pre-process is to transform each data instance into a high-dimensional sparse vector via one-hot or multi-hot encoding. 
A field is represented as a multi-hot encoding vector only when it is multivariate.
The data instance can be represented as 
\[\textbf{\textit{x}} = [\textbf{\textit{x}}_1, \textbf{\textit{x}}_2, \cdots, \textbf{\textit{x}}_n],\]
where $n$ is the number of fields and $\textbf{\textit{x}}_i$ is the one-hot or multi-hot encoding vector of the $i^{th}$ field. Then, a feature embedding layer is used to transform the encoding vector into a low-dimensional vector via  
\begin{equation}
    \textbf{\textit{e}}_i = V_i \textbf{\textit{x}}_i,
\end{equation}
where $V_i\in R^{d\times h_i}$ is the embedding matrix for the $i^{th}$ field, $h_i$ is vocabulary size of the $i^{th}$ field and $d$ is embedding size.


Then, the output of the  feature embedding layer is represented as the concatenation of multiple embedding vectors:
\[
\textbf{\textit{E}} = [\textbf{\textit{e}}_1, \textbf{\textit{e}}_2, ..., \textbf{\textit{e}}_n].
\]
 
In the traditional paradigm, all feature embeddings in various fields have identical embedding sizes. However, not all features in distinct fields are equally predictive, i.e., some fields are informative while the other fields are not. Therefore, allocating the same embedding size to all features may lose the information of high predictive features while wasting memory on non-predictive features. To tackle this issue, we will later present 
EDS component to search different embedding sizes (by pruning useless embedding dimensions) for various features in the different fields.

\textbf{Feature Interaction Layer.}
After transforming the features to a low-dimensional space,  the feature interactions can be modeled in such a space with the feature interaction layer. 
Various interaction functions (IFs) can be used to model distinctive interactive signals, such as inner product, outer product and kernel product:
%
%
\begin{equation}
\left\{
 \begin{aligned}
 f_{\texttt{INNER }}\left(\textbf{\textit{e}}_{i}, \textbf{\textit{e}}_{j}\right) &=\left \langle \textbf{\textit{e}}_{i}, \textbf{\textit{e}}_{j}\right\rangle\\
  f_{\texttt{OUTER }}\left(\textbf{\textit{e}}_{i}, \textbf{\textit{e}}_{j}\right) & = \textbf{\textit{e}}_{i} \otimes \textbf{\textit{e}}_{j}\\
 f_{\texttt{KERNEL}}\left(\textbf{\textit{e}}_{i}, \textbf{\textit{e}}_{j}\right) &=\left \langle \textbf{\textit{e}}_{i}, \textbf{\textit{e}}_{j}\right\rangle_{\phi}\\
 \end{aligned},
 \right.
 \end{equation}
where $\left \langle \cdot, \cdot\right\rangle$ is the inner
product, $\otimes$ is the outer product and $\left \langle \cdot, \cdot\right\rangle_{\phi}$ is the kernel product (with the kernel $\phi$). Specially, we further divide the kernel product into matrix kernel product, vector kernel product, scalar kernel product according to the shape of kernel $\phi$. 
 
Take the inner product as an example, we introduce how the feature interaction layer works.
The inner product of all the pair-wise feature interactions is calculated:
\begin{equation}
[\langle\textbf{\textit{e}}_{1}, \textbf{\textit{e}}_{2}\rangle, \langle\textbf{\textit{e}}_{1}, \textbf{\textit{e}}_{3}\rangle, ..., \langle\textbf{\textit{e}}_{n-1}, \textbf{\textit{e}}_{n}\rangle],
\end{equation}

In FM, the output of the feature interaction layer is: 
\begin{equation}
    l_{\texttt{FM}}=\langle \boldsymbol{w},\boldsymbol{x}\rangle +\sum_{i=1}^n\sum_{j>i}^n \langle \textbf{\textit{e}}_{i}, \textbf{\textit{e}}_{j}\rangle.
    \label{eqn:FM_inter1}
\end{equation}
 
However, the inner product operation may not be able to learn the interaction information of all the pairwise features. That is to say, various interaction functions (IFs) are needed when modeling different feature interactions, as also stated in~\cite{autoML_for_CF}. Taking this concern into consideration, we extend FM model with multiple IFs as
\begin{equation}
l_{\texttt{FM\_EXTEND}}=\langle \boldsymbol{w},\boldsymbol{x}\rangle +\sum_{i=1}^n\sum_{j>i}^n\sum_{k=1}^m f_k\left( \textbf{\textit{e}}_{i}, \textbf{\textit{e}}_{j}\right),
\label{eqn:FM_extend}
\end{equation}
where $f_k$ is a predefined IF (such as inner product, outer product, kernel product) and $m$ is the number of IFs.

In Equation~\ref{eqn:FM_extend}, all the feature interactions are assumed to be equally important and therefore are designed to contribute equally to the prediction. Whereas not all the feature interactions are equally predictive and useless interactions may even degrade the performance. 
Therefore, we propose the FIS component to select important feature interactions automatically and efficiently. 

In addition, Equation~\ref{eqn:FM_extend} assumes $k$ different IFs are all needed and contribute equally to model each feature interaction. As stated earlier, different IFs are needed when modeling individual feature interactions. To select suitable IFs for each feature interaction, we propose the IFS component, which will also be elaborated later. 

Besides pairwise feature interactions, we also target at identifying effective high-order feature interactions.  
Formally, we define $p^{th}$-order feature interaction (i.e., the combination of $p$ fields) as:
\begin{equation}
 \sum_{q\in \psi_p}f_k\left( \textbf{\textit{e}}_{q_1}, \textbf{\textit{e}}_{q_2}, \cdots, \textbf{\textit{e}}_{q_p} \right),  
\end{equation}
where $\psi_p$ contains all $p^{th}$-order interactions ($|\psi_p| = \mathcal{C}_{n}^{p}$) and $q$ is one such $p^{th}$-order interaction.
$f_k\left(\cdot\right)$ is the $k^{th}$ high-order IF (high-order IFs are expanded by $2^{nd}$-order IFs and the specific expansions are described in Section~\ref{sec:exp_implementation_detail}).


The complexity of feature interaction layer with $p^{th}$-order interaction is $O(n^p)$, exponential to the number of fields $n$, which makes searching high-order feature interactions in a brute-force manner unaffordable in practice. To tackle the efficiency issue, we propose a method to select high-order feature interactions with quadratic complexity, which will be illustrated in Section~\ref{sec:autofis}.

\textbf{MLP Layer.} 
MLP layer consists of several fully connected layers with activation functions, which learn the implicit non-linear relationship among features. The output of one such layer is
    \begin{equation}\label{eqn:MLP}
        \boldsymbol{a}^{(l+1)}=\mbox{relu}(W^{(l)}\boldsymbol{a}^{(l)}+\boldsymbol{b}^{(l)}),
    \end{equation}
where $\boldsymbol{a}^{(l)}, W^{(l)}, \boldsymbol{b}^{(l)}$ are the input, weight and bias of the $l^{th}$ layer, respectively. The activation function is $\mbox{relu}(z)=\max(0,z)$. $\boldsymbol{a}^{(0)}$ is the input to MLP layers and $\boldsymbol{a}^{(L)} = \texttt{MLP}(\boldsymbol{a}^{(0)})$, where $L$ is the depth of MLP layer. 

\textbf{Output Layer.}
 FM model has no MLP layer and connects the feature interaction layer with prediction layer directly:
\begin{equation}
    \hat{y}_{\texttt{FM}} = \mbox{sigmoid}( l_{\texttt{FM}} ) = \frac{1}{1 + \exp (-l_{\texttt{FM}})},
\end{equation}
where $\hat{y}_{\texttt{FM}} $ is the predicted CTR. DeepFM combines feature interaction layer and MLP layers in parallel as 
\begin{equation}
    \hat{y}_{\texttt{DeepFM}} = \mbox{sigmoid}(l_{\texttt{FM}}+ \texttt{MLP}(\boldsymbol{E})).
    \label{eqn:deepfm}
\end{equation}

While in IPNN, MLP layer is subsequent to feature interaction layer as 
\begin{equation}
 \hat{y}_{\texttt{IPNN}} = \mbox{sigmoid}(\texttt{MLP}([\boldsymbol{E},l_{\texttt{FM}}])).   
\end{equation}

Note that the MLP layer of IPNN can also serve as a re-weighting of the different feature interactions to capture their relative importance. This is also the reason that IPNN has a higher capacity than FM and DeepFM. However, with the IPNN formulation, one cannot retrieve the exact value corresponding to the relative contribution of each feature interaction. Therefore, the useless feature interactions in IPNN can be neither identified nor dropped, which brings extra noise and computation cost to the model. 

\textbf{Objective Function.} 
 FM, DeepFM, and IPNN share the same objective function,  i.e., to minimize the cross-entropy of predicted values $\hat{y}$ and the labels $y$ as 
\begin{equation}
    \mathcal{L}(y,\hat{y}) = -y\mbox{log}\hat{y}- (1-y)\mbox{log}(1-\hat{y}),
    \label{eqn:loss}
\end{equation}
where $y\in\{0,1\}$ is the label and $\hat{y} \in [0,1]$ is the predicted probability of $y=1$. 

\subsection{Feature Interaction Search (FIS)}\label{sec:autofis}

In this section, we present one of the three core components, namely FIS component, which is performed on the feature interaction layer of any factorization model to automatically identify different orders of essential feature interactions.

\begin{figure}[h]
    \centering
    \vspace{-0.1em}
    \includegraphics[width=0.38\textwidth]{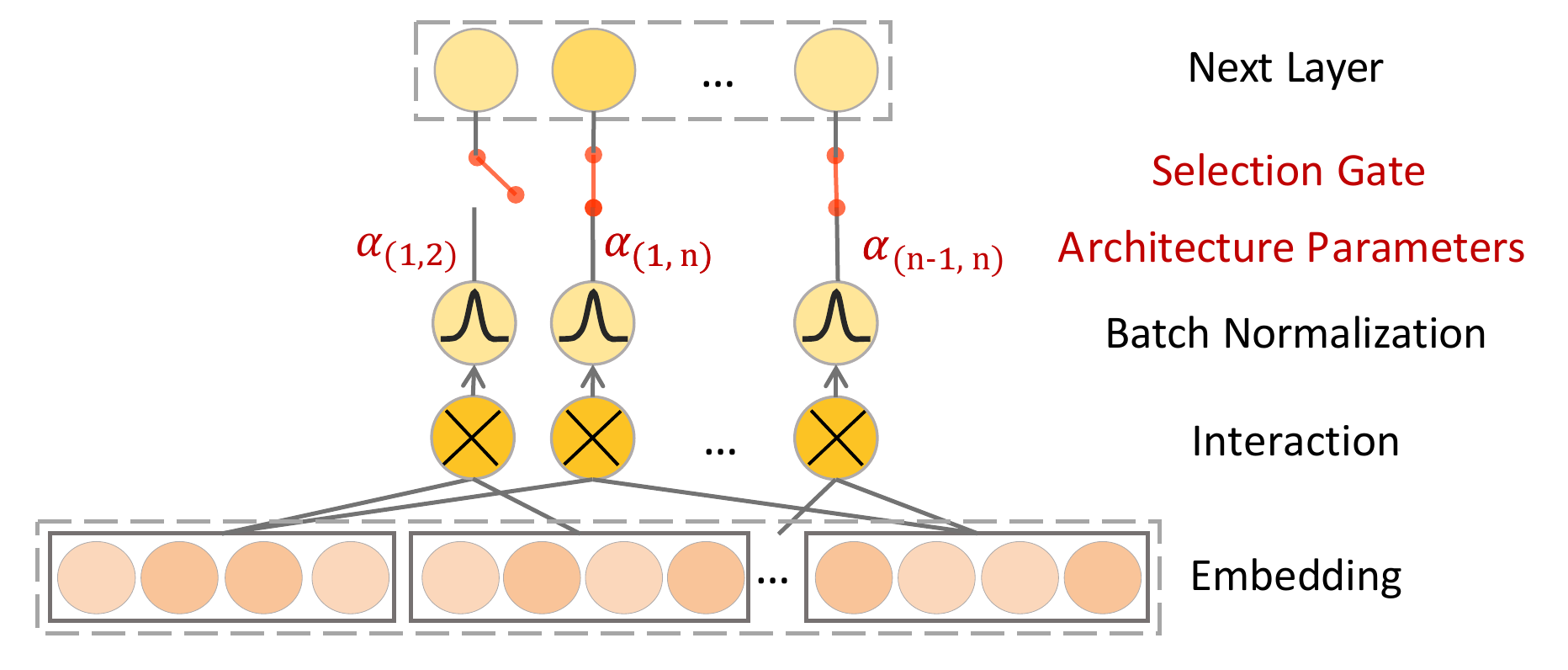}
    \vspace{-1.2em}
    \caption{ Overview of FIS component.}
    \vspace{-1.3em}
    \label{fig:autoFIS}
\end{figure}


The overview of the FIS component is illustrated in Figure~\ref{fig:autoFIS}. To ease the presentation of FIS component, we introduce the \emph{gate} operation to control whether to select a feature interaction, i.e., an open gate corresponds to selecting this feature interaction, while a closed gate results in dropping this interaction. The total number of gates corresponding to all the $2^{nd}$-order feature interactions is $\mathcal{C}_{n}^{2}$. It is very challenging to find the optimal set of open gates in a brute-force way, as we face an incredibly huge ($2^{\mathcal{C}_{n}^{2}}$) space to search. In this work, we approach the problem from a different viewpoint. Instead of searching over a discrete set of open gates, we relax the choices to be continuous by introducing architecture parameters $\boldsymbol{\alpha}$, so that the relative importance of each feature interaction can be learned by gradient descent. 

This architecture selection scheme based on gradient learning is inspired by DARTS \cite{liu2018darts}, where the objective is to select one operation from a set of candidate operations in convolutional neural network (CNN) architecture. 
To be specific, we reformulate the interaction layer in factorization models (shown in Equation~\ref{eqn:FM_inter1}) as
\begin{equation}
    l_{\texttt{FIS}}  = \langle \boldsymbol{w},\boldsymbol{x}\rangle +\sum_{i=1}^n\sum_{j>i}^n \alpha_{(i,j)}\langle\textbf{\textit{e}}_{i}, \textbf{\textit{e}}_{j}\rangle,
    \label{eqn:FM_autointer}
\end{equation}
where $\boldsymbol{\alpha}=\{\alpha_{(1,2)}, \cdots, \alpha_{(n-1,n)}\}$ are the architecture parameters and their value $\alpha_{(i,j)}$ can represent the relative contribution of each feature interaction to the final prediction. Then, we can decide the gate status of each feature interaction by setting those unimportant ones (i.e., with zero $\alpha_{(i,j)}$ values) closed.

After the search, some unimportant interactions are thrown away automatically according to the architecture parameters $\boldsymbol{\alpha}$ and the new model with remained feature interactions can be re-trained. 

\textbf{High-order Feature Interaction Search.}
Note that, Equation~\ref{eqn:FM_autointer} only formulates the case of selecting important $2^{nd}$-order feature interactions. 
Besides the $2^{nd}$-order, FIS component also aims to search high-order feature interactions. 
The number of all $p^{th}$-order feature interaction is $\mathcal{C}_{n}^{p}$, which is exponential to the number of fields $n$. Therefore, it is not practical to identify effective high-order feature interactions in a similar way as for $2^{nd}$-order feature interactions (by enumerating all the feature interactions first).

To search effective high-order feature interactions efficient, we propose a two-step process to select $p^{th}$-order feature interactions from the selected $(p-1)^{th}$-order interactions, as follows.

\begin{itemize}[leftmargin=12pt]
	\item Step 1: Exploit $(p-1)^{th}$-order and first-order feature interaction set to generate a preliminary candidate pool $\mathcal{M}_p^{tmp}$ for potential effective $p^{th}$-order interactions.
	\item Step 2: Perform FIS component with architecture parameters to select effective $p^{th}$-order feature interactions $\mathcal{M}_p$, from $\mathcal{M}_p^{tmp}$ generated by Step 1.
	\end{itemize}
\vspace{-0.5em}
\begin{algorithm}[!htb]
	\caption{ High-order Feature Interaction Search}
	\label{alg:high-order selection}	
	\begin{algorithmic}[1]
		\REQUIRE
		field number $n$, the highest order $P$, top number $k$ for pool selection 
		\ENSURE
		the selected interactions $\mathcal{M}_1, \mathcal{M}_2, \mathcal{M}_3, \cdots, \mathcal{M}_P$
		\STATE{$\mathcal{M}_1 \leftarrow \{ (\textbf{\textit{e}}_1),(\textbf{\textit{e}}_2), \cdots ,(\textbf{\textit{e}}_n)\}$}
		\STATE{$\mathcal{M}_1^{top} \leftarrow \mathcal{M}_1$}
		\FOR{$p\leftarrow 2$ to $P$}
		\STATE{$\mathcal{M}_p^{tmp} \leftarrow$ Combine($\mathcal{M}_{p-1}^{top},\mathcal{M}_1$)}
		\STATE{$\mathcal{M}_p, \boldsymbol{\alpha_p} \leftarrow$ FIS($\mathcal{M}_p^{tmp}$)}
		\STATE{Sort $\mathcal{M}_p$ by $\boldsymbol{\alpha_p}$}
		\STATE{$\mathcal{M}_p^{top} \leftarrow Top_k(\mathcal{M}_p)$}
		\ENDFOR
		\RETURN{ $\mathcal{M}_1, \mathcal{M}_2, \mathcal{M}_3, \cdots, \mathcal{M}_P$}
	\end{algorithmic}
\end{algorithm}
\vspace{-0.5em}
This two-step search process is summarized in Algorithm~\ref{alg:high-order selection}. We elucidate the core part of this algorithm, i.e., \emph{line 4} to \emph{line 7}. 
For \emph{line 4}, we combine top important ${(p-1)}^{th}$-order feature interaction set $\mathcal{M}_{p-1}^{top}$ and first-order feature interaction set $\mathcal{M}_1$ (that is each feature) in a Cartesian product manner to get the candidate pool $\mathcal{M}_p^{tmp}$ for ${p}^{th}$-order. 
For \emph{line 5}, we leverage the FIS component to learn the relative importance of each $p^{th}$-order interaction by the architecture parameters.
More specifically, we abandon the feature interactions with zero $\alpha$ value in $\mathcal{M}_p^{tmp}$ and retain the rest to get the final selected $p^{th}$-order interaction set $\mathcal{M}_p$ with their corresponding $\boldsymbol{\alpha_p}$.
For \emph{line 6-7}, preparing for the next round, we employ $\boldsymbol{\alpha_p}$ to sort $\mathcal{M}_p$ and get the top-$k$ ($k\leq n$) important $p^{th}$-order feature interaction to generate ${(p+1)}^{th}$-order feature interaction set.

\textbf{Complexity Analysis.} To select effective $p^{th}$-order feature interactions, there exists at most $n$ feature interactions in $\mathcal{M}_{p-1}^{top}$, therefore $\mathcal{M}_p^{tmp}$ (which combining $\mathcal{M}_{p-1}^{top}$ and $\mathcal{M}_1$) has at most $O(n^2)$ $p^{th}$-order feature interactions. Both time and space complexity of feature interaction selection for each order by our method is $O(n^2)$, instead of $O(n^{p})$ complexity by simple enumeration. 

\subsection{Interaction Function Search (IFS)}

FIS component selects important feature interactions which are modeled with the same specific interaction function (IF). However, as stated earlier, not all the feature interactions can be modeled with the same IF. That is to say, individual feature interactions may need different IFs. Due to this reason, the second core component, IFS, is proposed to select appropriate IFs for each important feature interaction.

\begin{figure}[htbp]
	\centering
	\vspace{-1.4em}
	\includegraphics[width=0.46\textwidth]{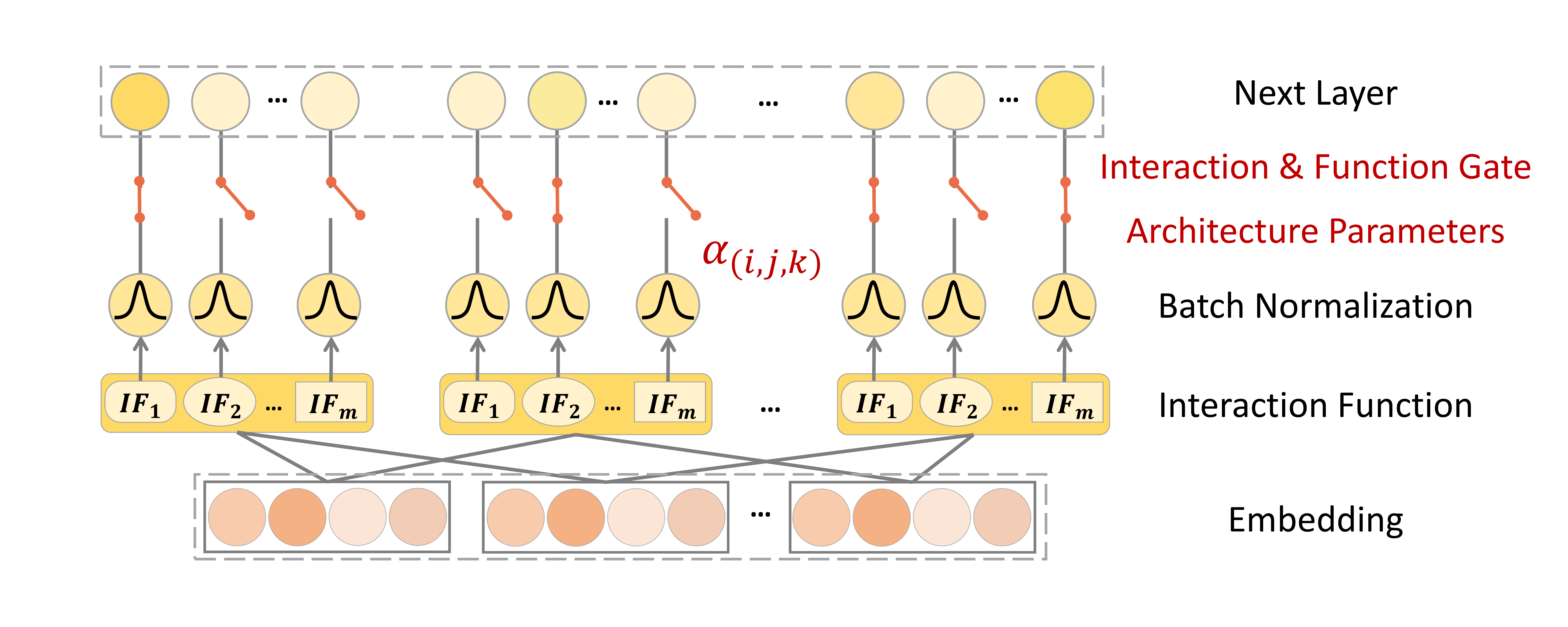}	
	\vspace{-1.9em}
	\caption{ The architecture of IFS component.}
	\vspace{-0.3em}
	\label{fig:AutoFIFS}
\end{figure}

Similar to FIS, IFS component also leverages the \emph{gate} operation to control the IF selection. For each feature interaction, one or more IFs from $m$ candidate IFs are selected to learn the relationship of the features. As a special case, a feature interaction is abandoned if none of $m$ IFs is selected.
The architecture parameters $\boldsymbol{\alpha'}$ are introduced to get the relative importance of each interaction-IF pair, which can be learned by gradient descent (in a similar way as in FIS).
The overview of IFS component is presented in Figure~\ref{fig:AutoFIFS}. 

We reformulate the interaction layer in the extend factorization models (shown in Equation~\ref{eqn:FM_extend}) as 
\begin{equation}
    l_{\texttt{IFS}}  = \langle \boldsymbol{w},\boldsymbol{x}\rangle +\sum_{i=1}^n\sum_{j>i}^n\sum_{k=1}^m \alpha'_{(i,j,k)}f_k(\textbf{\textit{e}}_{i}, \textbf{\textit{e}}_{j}),
    \label{eqn:AutoFIFS}
\end{equation}
where $\boldsymbol{\alpha'}=\{\alpha_{(i,j,k)}'\}$ are the architecture parameters. $\alpha'_{(i,j,k)}$ corresponding to the $k^{th}$ IF for feature interaction $(i, j)$. Those unimportant interaction-IF pairs (i.e., with zero $\alpha'_{(i,j, k)}$ values) will be thrown away automatically.

\textbf{Function-wise Embeddings.}
In Equation~\ref{eqn:AutoFIFS}, feature interaction modeling leverages shared embedding (i.e., each feature shares the same embedding for different IFs) to learn the latent effect over different IFs. 
However, the learning of shared feature embedding becomes difficult as one such embedding receives different gradient signals from individual learning spaces, where each learning space corresponds to an IF. Such different gradient signals cause learning conflict and lead to sub-optimal performance, as also observed in~\cite{OENN,CAN}.

To avoid such learning conflict among different IFs, IFS proposes to utilize the function-wise embeddings (FWEs), i.e., multiple embeddings are learned for each feature, where each of such embedding corresponds to one IF.
With a specific IF, the corresponding embedding vectors are used to model the interactive correlations. Equipped with FWE, Equation~\ref{eqn:AutoFIFS} is updated as:
\begin{equation}
     l_{\texttt{IFS}}^{\texttt{FWE}}  = \langle \boldsymbol{w},\boldsymbol{x}\rangle +\sum_{i=1}^n\sum_{j>i}^n\sum_{k=1}^m \alpha'_{(i,j,k)}f_k(\textbf{\textit{e}}_{ik}, \textbf{\textit{e}}_{jk}),
    \label{eqn:AutoFIFS_operation_wised}
\end{equation}
where $\textbf{\textit{e}}_{ik}$ represents the embedding of the $i^{th}$ feature for the $k^{th}$ IF. 
In this way, each embedding only needs to learn the latent effect in one mapping space of a specific IF, such that learning conflict is avoided. 
In Section~\ref{sec:exp_ablation_study}, we will show the superiority of function-wise embeddings over shared embedding.
However, function-wise embeddings introduce more parameters, which leads to much more space cost. To reduce the space complexity, the third core component, EDS is proposed, which will be presented in Section~\ref{sec:autoembgate}.

After the architecture parameters $\boldsymbol{\alpha'}$ are learned, appropriate IFs are chosen for each feature interaction automatically. More specifically, zero value of $\alpha'_{(i,j,k)}$ indicates the interaction $(i,j)$ with $k^{th}$ IF is not effective and thus can be dropped. The other interaction-IF pairs with non-zero $\alpha'_{(i,j,k)}$ values are retained.

\subsection{Embedding Dimension Search (EDS)}\label{sec:autoembgate}

Although function-wise embeddings (FWEs) improve the performance, the model size is significantly larger than the case of allocating the same embedding size (as shared embedding) to all features. To deal with the excessive parameters, the third component, EDS, is proposed to optimize the model size without sacrificing performance by assigning low embedding size to non-predictive features, which is achieved by pruning redundant embedding dimensions for each feature.

\begin{figure}[htbp]
	\centering
	\vspace{-0.6em}
	\includegraphics[width=0.43\textwidth]{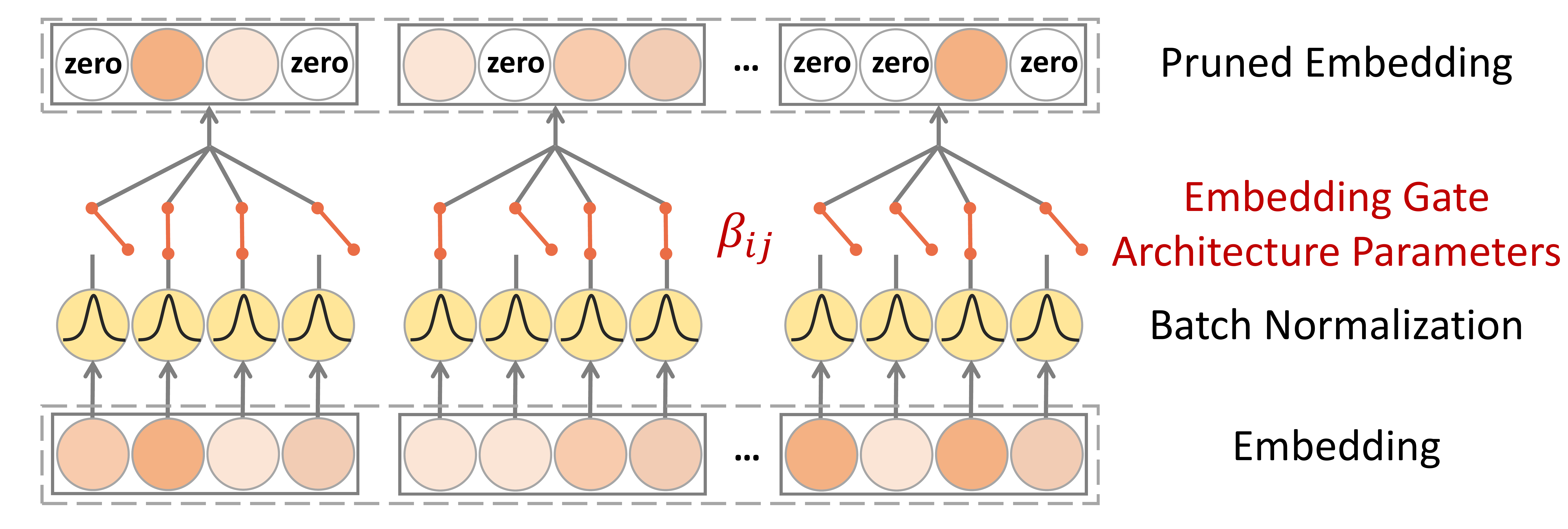}	
	\vspace{-0.9em}
	\caption{ The architecture of EDS component.}
	\vspace{-0.8em}
	\label{fig:search_embed}
\end{figure}

Similarly, a \emph{gate} operation is introduced to determine whether the corresponding embedding dimension is selected for a field. 
The total number of the gates is $n\times d$, where $n$ is the number of the fields and $d$ is the largest embedding size. Here we make use of architecture parameters $\boldsymbol{\beta}\in \mathbb{R}^{n\times d}$ to turn the embedding dimension search problem into a continuous process which can be solved by gradient descent. The overview of EDS component is shown in Figure~\ref{fig:search_embed}. Note that FWEs in the same field share the same set of architecture parameters.

During the embedding dimension search, each embedding dimension multiplies to the embedding architecture parameter $\boldsymbol{\beta}$ of the corresponding field, as
\begin{equation}
     \textbf{\textit{e}}_i' = \boldsymbol{\beta_i}\cdot \textbf{\textit{e}}_i,
\end{equation}
where $\boldsymbol{\beta_i}$ is the embedding architecture parameter of the $i^{th}$ field and it represents the relative importance among different dimensions in the $i^{th}$ field. 

The feature embedding layer concatenates the embedding vector $\textbf{\textit{e}}_i'$ of each feature, as
\[
\textbf{\textit{E}}' = [\textbf{\textit{e}}_1', \textbf{\textit{e}}_2', ..., \textbf{\textit{e}}_n'].
\]

Then $\textbf{\textit{E}}'$ feeds to the feature interaction layer for capturing interactive signals.

After the search, the embedding size of each feature field has been determined automatically according to $\boldsymbol{\beta}^*$. More specifically, EDS component prunes redundant dimensions (i.e., with zero $\beta_{ij}$ value) and reserves necessary dimensions with their positions (since the positions also reflect the effect on various feature interactions). The embedding size for each field is
\begin{equation}
d_i = \sum_{j=1}^{d} \boldsymbol{\mathds{1}}\left[ {\beta}_{ij}\neq 0 \right].
\end{equation}

Besides, the position set of the retained embedding dimensions for the $i^{th}$ field is 
\begin{equation}
\phi_i = \{j\,|\,{{\beta}_{ij}\neq 0}\ , \ j\in\left[1,\cdots, d\right] \}.
\end{equation}
 
To derive new embeddings of adaptive sizes, we design a function $Map_i\left(\cdot\right)$ for the $i^{th}$ field that 
maps from the position set of retained dimension $\phi_i$ to the new embedding position set $\{1, 2, \cdots, d_i\}$ in order as
\begin{equation}
    Map_i:  \phi_i \longrightarrow \{1, 2, \cdots, d_i\},
\end{equation}
where $d_i$ is the new embedding size of the $i^{th}$ fields.
Then we can re-train the model with length-adaptive embedding sizes. For each feature $e_i$ in the $i^{th}$ field, instead of initializing it with embedding size $d$, we initialize it with a shorter embedding size $d_i$.

As some IFs require embedding sizes of features to be the same (such as inner product), in such cases, we can trivially utilize $Map_i$ to transform embeddings with size $d_i$ to ones with size $d$, by setting the dimensions that are not in the domain of $Map_i$ with zero values. Note that such transformation with $Map_i$ introduces no extra parameters therefore does not increase the space complexity. 





\subsection{Automatic Interaction Machine (AIM)}
\label{sec:autofifs_emb}


To summarize, FIS component searches for feature interactions; IFS component searches for appropriate IF for each feature interaction; and EDS component searches for proper embedding size for each feature. To fully exploit the advantages of these three components, we build the AIM framework by combining them, to select significant feature interactions, appropriate IFs and necessary embedding dimensions automatically in a unified framework.

\begin{figure}[htbp]
	\centering
	\vspace{-0.9em}
	\includegraphics[width=0.48\textwidth]{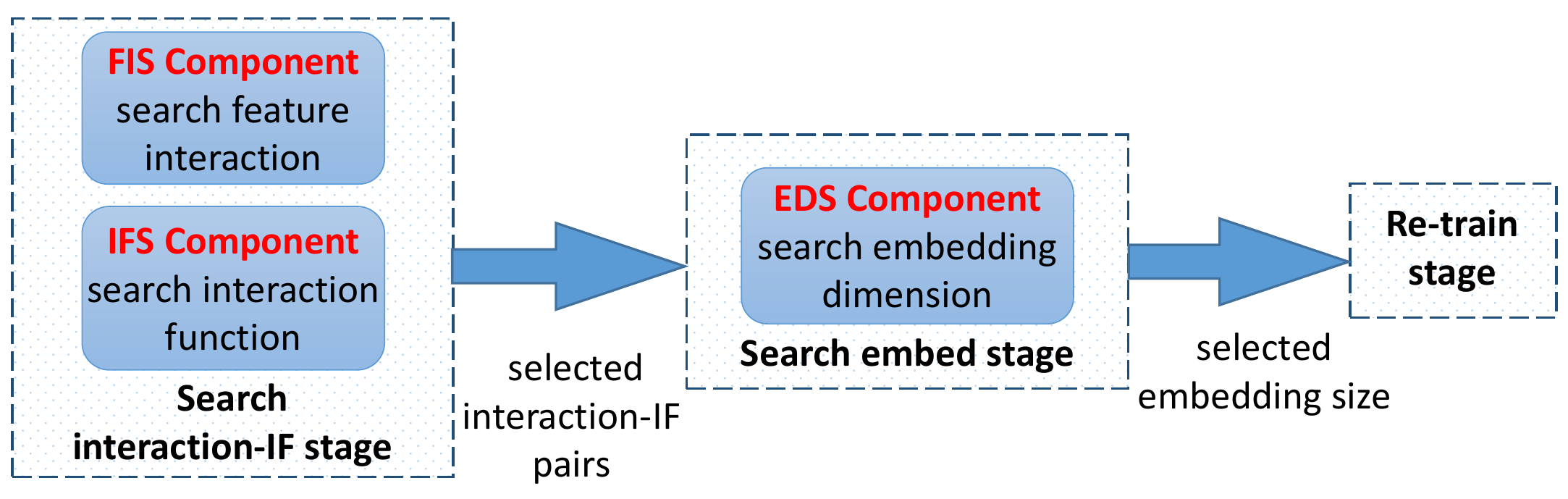}	
	\vspace{-1.2em}
	\caption{ The flow chart of AIM.}
	\vspace{-0.5em}
	\label{fig:aim_flow}
\end{figure}

Specifically, AIM has the following three stages as shown in Figure~\ref{fig:aim_flow}:

\begin{itemize}[leftmargin=12pt]
	\item \textbf{Search interaction-IF stage.} Train the model and then use FIS component and IFS component to select useful interactions and appropriate IFs.
	\item \textbf{Search embed stage.} Train the model with selected interaction-IF pairs in the first stage, and then use EDS component to select a suitable embedding size for each field, by discarding useless embedding dimensions.
	\item \textbf{Re-train stage.} Re-train the model with selected interaction-IF pairs and learned embedding sizes in the first two stages.
\end{itemize}

Note that FIS component and IFS component can be trained together, however, it is inappropriate to combine the training process of them with EDS component due to the coupling problem between architecture parameters. Therefore, we have to divide the search process into Search interaction-IF stage and Search embed stage.

\textbf{Transferability.} 
The chosen Interaction-IF pairs by \emph{gate} architecture learned from a simple model could be transferred to the state-of-the-art models to boost their performance. We will verify this transferability in Section~\ref{sec:exp_transfer}. 

\subsection{Training Details}
\label{sec:learning}
In this subsection, we present some training details in AIM, i.e., batch normalization and GRDA Optimizer. We take FIS component as an illustration example and similar training procedures are performed in IFS and EDS.

\textbf{Batch Normalization.}
From the viewpoint of the overall neural network, the contribution of a feature interaction is measured by $\alpha_{(i,j)}\cdot \langle \textbf{\textit{e}}_i, \textbf{\textit{e}}_j\rangle$ (in Equation~\ref{eqn:FM_autointer}). Exactly the same contribution can be achieved by re-scaling this term as  $(\frac{\alpha_{(i,j)}}{\eta})\cdot (\eta\cdot \langle \textbf{\textit{e}}_i, \textbf{\textit{e}}_j\rangle)$, where $\eta$ is a real number.

Since the value of $\langle \textbf{\textit{e}}_i, \textbf{\textit{e}}_j\rangle$ is jointly learned with $\alpha_{(i,j)}$, the coupling of their scale would lead to unstable estimation of $\alpha_{(i,j)}$, such that $\alpha_{(i,j)}$ can no longer represent the relative importance of $\langle \textbf{\textit{e}}_i, \textbf{\textit{e}}_j\rangle$. 
To solve this problem, we apply Batch Normalization (BN)~\cite{bn} on $\langle \textbf{\textit{e}}_i, \textbf{\textit{e}}_j\rangle$ to eliminate its scale issue. BN has been adopted by training deep neural networks as a standard approach to achieve fast convergence and better performance. The way that BN normalizes values gives an efficient yet effective way to solve the coupling problem of $\alpha_{(i,j)}$ and $\langle \textbf{\textit{e}}_i, \textbf{\textit{e}}_j\rangle$. 

The original BN normalizes the activated output with statistics information of a mini-batch. Specifically, 
\begin{equation}
\hat{\boldsymbol{z}} = \frac{\boldsymbol{z}_{\texttt{IN}}-\boldsymbol{\mu}_{\mathcal{B}}}{\sqrt{
\sigma_{\mathcal{B}}^{2}+\epsilon}} \hspace{10pt}\text{and}\hspace{10pt} \boldsymbol{z}_{\texttt{OUT}} = \theta \cdot \hat{\boldsymbol{z}} + \boldsymbol{\delta},    
\end{equation}
where $\boldsymbol{z}_{\texttt{IN}}$, $\hat{\boldsymbol{z}}$ and $\boldsymbol{z}_{\texttt{OUT}}$ are input, normalized and output values of BN; $\boldsymbol{\mu}_{\mathcal{B}}$ and $\sigma_{\mathcal{B}}$ are the mean and standard deviation values of $\boldsymbol{z}_{\texttt{IN}}$ over a mini-batch $\mathcal{B}$; $\theta$ and $\boldsymbol{\delta}$ are trainable \emph{scale} and \emph{shift} parameters of BN; $\epsilon$ is a constant for numerical stability.

To get stable estimation of $\alpha_{(i,j)}$, we set the \emph{scale} and \emph{shift} parameters to be 1 and 0 respectively. The BN operation on each feature interaction $\langle \textbf{\textit{e}}_i,\textbf{\textit{e}}_j\rangle$ is calculated as 
\begin{equation}
\langle \textbf{\textit{e}}_i,\textbf{\textit{e}}_j\rangle_{\texttt{BN}} = \frac{\langle \textbf{\textit{e}}_i,\textbf{\textit{e}}_j\rangle - \mu_{\mathcal{B}}(\langle \textbf{\textit{e}}_i,\textbf{\textit{e}}_j\rangle)}{\sqrt{
\sigma_{\mathcal{B}}^{2}(\langle \textbf{\textit{e}}_i,\textbf{\textit{e}}_j\rangle)+\epsilon}},
\end{equation}
where $\mu_{\mathcal{B}}$ and $\sigma_{\mathcal{B}}$ are the mean and standard deviation of $\langle \textbf{\textit{e}}_i,\textbf{\textit{e}}_j\rangle$ in mini-batch $\mathcal{B}$.

\textbf{GRDA Optimizer}.
 Generalized  regularized dual averaging (GRDA) optimizer~\cite{chao2019generalization} is aimed to get a sparse  deep neural network. 
To update $\alpha$ at each gradient step $t$ with data $Z_{t}$ we use the following equation:
\begin{equation}
\footnotesize
    \alpha_{t+1} = \arg \min \limits_\alpha \{\alpha^T(-\alpha_0+\gamma \sum_{i=0}^t \nabla  L(\alpha_t;Z_{i+1}))+g(t,\gamma)\|\alpha\|_1+\frac{1}{2}\|\alpha\|_2^2\}
\end{equation}
where $g(t,\gamma)=c \gamma^{1/2}(t\gamma)^u$, and $\gamma$ is  the learning rate,  $c$ and $\mu$ are adjustable hyper-parameters to trade-off between accuracy and sparsity.

With the GRDA optimizer, the important feature interactions (or functions, embedding dimensions) will be retained, and the unnecessary ones will be abandoned according to the architecture parameters (the unimportant ones will be learned to zero). 
In this way, GRDA optimizer adaptively determines the remained ones, avoiding artificially setting hyper-parameters for deciding how many feature interactions (or functions, embedding dimensions) to be remained.

%% file: exp.tex
\section{Experiments}
\label{sec:exp}

In this section, we conduct extensive offline experiments\footnote{\scriptsize Repeatable code of all the experimental study and all hyper-parameters are available at https://github.com/zhuchenxv/AIM \label{code} } on two public benchmark datasets and a private dataset, as well as an online A/B test in the recommendation service of a mainstream app market, to evaluate the effectiveness of Automatic Interaction Machine (AIM). In particular, we answer the following research questions:


\begin{itemize}[leftmargin = 12 pt]
	\item \textbf{RQ1}: Can AIM outperform the state-of-the-art models with the selected interaction-IF pairs?
	\item \textbf{RQ2}: How do different components of AIM (e.g., FIS, IFS, EDS) contribute to the performance? Are the interactions, IFs and embedding dimensions selected by these components really important and valuable?
	\item \textbf{RQ3}: How about the space and time complexity of AIM compared with other models?
	\item \textbf{RQ4}: Can the interaction-IF pairs selected from AIM be transferred to other models for boosting their performance?
	\item \textbf{RQ5}: Can AIM improve the performance of an existing model in a live recommender system?
\end{itemize}

\subsection{Datasets}

Experiments are conducted on the following two public datasets (Avazu and Criteo) and one private dataset (Huawei), whose statistics are summarized in Table~\ref{tab:dataset}. We follow the existing works~\cite{pnn,pin,autofis,autofeature,autogroup} to process Avazu and Criteo datasets.

\textbf{Avazu\footnote{{\scriptsize http://www.kaggle.com/c/avazu-ctr-prediction}}}: Avazu was released in the CTR prediction contest on Kaggle. $80\%$ of randomly shuffled data is allotted to training and validation with $20\%$ for testing. 

\textbf{Criteo\footnote{{\scriptsize http://labs.criteo.com/downloads/download-terabyte-click-logs/}}}: Criteo contains one month of click logs with billions of data samples. We select ``data 6-12'' as training and validation set while select ``day-13'' for evaluation. 

\textbf{Huawei}: The industrial dataset is sampled and collected from an app recommendation scenario of Huawei AppGallery for a week. The dataset contains 10 feature fields, including user behavior (user click list, \textit{etc.}), app information (id, category, \textit{etc.}), and context information (time, \textit{etc.}).


\begin{table}[h]
	\vspace{-0.6em}
	\caption{ Dataset Statistics}
	\vspace{-1em}
	\label{tab:dataset}
	\centering
	\resizebox{0.35\textwidth}{!}{
		\begin{tabular}{ccccc}
			\toprule
			Dataset & \#instances & \#features & \#fields & pos ratio \\ \midrule 
			Avazu & $4\times 10^{7}$ & $6\times 10^{5}$ & 24 & 0.17 \\
			Criteo & $1 \times 10^{8}$ & $1\times 10^{6}$ & 39 & 0.50 \\
			Huawei & $3\times 10^{8}$ & $1\times 10^{5}$ & 10 & 0.07 \\\bottomrule
		\end{tabular}
	}
	\vspace{-0.6em}
\end{table}

\begin{table*}[t]
	\centering
	\vspace{-1.0em}
	\caption{ Benchmark performance.``Rel. Impr'' is the relative AUC improvement over FM model.}
	\vspace{-1.1em}
	\label{tab:performance_overall_final}
	\resizebox{0.8\textwidth}{!}{
	\begin{threeparttable}[t]{
	\begin{tabular}{l|c|ccc|ccc|ccc}
		\toprule
		\multicolumn{1}{c|}{\multirow{2}{*}{Category}}&
		\multicolumn{1}{c|}{\multirow{2}{*}{Model}}& \multicolumn{3}{c|}{Avazu} & \multicolumn{3}{c|}{Criteo}& \multicolumn{3}{c}{Huawei}\\
		\multicolumn{1}{c|}{}& & AUC & log loss& Rel. Impr. & AUC & log loss & Rel. Impr. & AUC & log loss & Rel. Impr. \\ 
		\midrule 
		\multicolumn{1}{c|}{\multirow{6}{*}{		\tabincell{c}{Factorization Model}
		}}&\multicolumn{1}{c|}{FM}  & 0.7793 & 0.3805 & 0 & 0.7909  &0.5500 &  0  & 0.8168 & 0.1939  & 0\\ 
		&\multicolumn{1}{c|}{FwFM}  & 0.7822 & 0.3784 & 0.37\% & 0.7948  &0.5475 &   0.49\% & 0.8268 &  0.1907  & 1.22\% \\ 
		&\multicolumn{1}{c|}{AFM} &  0.7806 &0.3794  & 0.17\% & 0.7913 & 0.5517 &  0.05\% & 0.8195 &  0.1934 & 0.33\% \\
		&\multicolumn{1}{c|}{FFM}  & 0.7831 & 0.3781 & 0.49\% & 0.7980  &0.5438 & 0.90\%  & 0.8279 & 0.1900  &  1.36\% \\
		&\multicolumn{1}{c|}{DeepFM}  &  0.7836 & 0.3776  & 0.55\%&  0.7991 & 0.5423  & 1.04\%  & 0.8338 & 0.1878  & 2.08\% \\
		&\multicolumn{1}{c|}{IPNN} & 0.7868 & 0.3756 & 0.96\%  & 0.8013 &  0.5401 &  1.31\% & 0.8363 &  0.1867  & 2.39\% \\
		\midrule
		\multicolumn{1}{c|}{\multirow{3}{*}{
		\tabincell{c}{Other Interaction Model}
		}}&\multicolumn{1}{c|}{Fi-GNN} & 0.7853  & 0.3767  &  0.77\% &  0.8003 & 0.5410   & 1.19\% &  0.8376 & 0.1863 & 2.55\% \\
		&\multicolumn{1}{c|}{AutoInt} &  0.7847 & 0.3770 & 0.69\%  &  0.8001  &  0.5413  & 1.16\% & 0.8371  & 0.1865 & 2.49\%  \\
		&\multicolumn{1}{c|}{xDeepFM} & 0.7855 & 0.3766 & 0.80\%  & 0.8006  & 0.5408    & 1.23\%  & 0.8368 & 0.1865  & 2.45\% \\
		\midrule
		\multicolumn{1}{c|}{\multirow{3}{*}{
		\tabincell{c}{AutoML-based Model}
		}}&\multicolumn{1}{c|}{AutoFeature} & 0.7904  & 0.3737 & 1.42\%  & 0.8023  & 0.5390  &  1.44\% & 0.8365  &  0.1861 & 2.41\%\\
		&\multicolumn{1}{c|}{AutoGroup} & 0.7909 & 0.3732 &  1.49\%  & 0.8026 & 0.5386  & 1.48\%  &  0.8360  & 0.1860  &2.35\% \\
		&\multicolumn{1}{c|}{AutoFIS}  &  0.7883 & 0.3748  & 1.15\%  &   0.8012 & 0.5402  &  1.30\% & 0.8373  & 0.1864  &2.51\%\\
 		\midrule
		\multicolumn{2}{c|}{\textbf{AIM}}  & \textbf{0.7920*} & \textbf{0.3727*}  &  1.63\% & \textbf{0.8030*}  & \textbf{0.5379*}  & 1.53\% &  \textbf{0.8385*} &  \textbf{0.1858*}    &2.66\%\\
		\bottomrule
	\end{tabular}
	\begin{tablenotes}
		\item \scriptsize $*$ denotes statistically significant improvement (measured by t-test with p-value$<$0.005) over  baselines with same order.
	\end{tablenotes}
}
\end{threeparttable}
}
\vspace{-1.0em}
\end{table*}

\subsection{Experimental Settings}

\subsubsection{Baselines and Evaluation Metrics}

We compare our propoed \textbf{AIM} with representative factorization models (i.e., FM~\cite{fm}, FwFM~\cite{fwfm}, AFM~\cite{afm}, FFM~\cite{ffm}, DeepFM~\cite{deepfm}, IPNN~\cite{pin}), AutoML-based model (AutoFeature~\cite{autofeature}, AutoGroup~\cite{autogroup} and AutoFIS~\cite{autofis}) and some other interaction model (xDeepFM~\cite{xdeepfm}, FiGNN~\cite{fignn}, AutoInt~\cite{autoint}). Note that AutoFIS use DeepFM as base model in our experiment.

We use the common evaluation metrics for CTR prediction: \textbf{AUC} and \textbf{Log loss}. All the experiments are repeated five times by changing the random seeds. The two-tailed unpaired $t$-test is performed to detect significant differences between our model and the best baseline.

\subsubsection{Parameter Settings}

In AIM model, we set the embedding size $d=40$ for Avazu dataset and $d=20$ for Criteo dataset. The MLP structure in two datasets are both $[700\times 5, 1]$. With regard to GRDA parameters $c$ and $mu$, we set $c=0.05, mu=0.6$ for Avazu dataset and $c=0.005, mu=0.9$ for Criteo dataset. The more detailed parameters for AIM model and the hyper-parameters for other models can be shown in our code link\textsuperscript{\ref{code}}.


\subsubsection{Implementation Details}\label{sec:exp_implementation_detail}

In the \emph{search interaction-IF stage}, we first train the model with $\boldsymbol{\alpha}'$ to seek important interaction-IF pairs. Then in the \emph{search embed stage}, we train the model with $\boldsymbol{\beta}$ to search embedding dimensions. Finally, we re-train the model with the selected components.
We consider $4$ different common-used IFs, containing inner product, outer product, vector kernel product and scalar kernel product. The output of outer product is a $d\times d$ vector ($d$ is the embedding size), and we use a linear layer to transform it into a single output. To reduce the complexity of outer product, we utilize $(\textbf{\textit{w}}_1^T\textbf{\textit{e}}_i)(\textbf{\textit{w}}_2^T \textbf{\textit{e}}_j) = \textbf{\textit{w}}_1^T\textbf{\textit{e}}_i \textbf{\textit{e}}_j^T \textbf{\textit{w}}_2$ to approximate outer product. Here $\textbf{\textit{e}}_i$, $\textbf{\textit{e}}_j$ are the embeddings and $\textbf{\textit{w}}_1$, $\textbf{\textit{w}}_2$ are learnable parameters.

For high-order interaction implementation such as $p^{th}$-order interaction, inner product, vector kernel product and scalar kernel product are easily to be extended by element-wise multiplication with $p$ vectors. To expand the outer product into $p^{th}$-order, we employ $(\textbf{\textit{w}}_1^T\textbf{\textit{e}}_{q_1})(\textbf{\textit{w}}_2^T \textbf{\textit{e}}_{q_2})\cdots (\textbf{\textit{w}}_p^T \textbf{\textit{e}}_{q_p})$ to calculate it ($q\in \psi_p$ where $\psi_p$ contains all $p^{th}$-order interactions).

To implement high-order interaction selection, we manipulate top $k$ (set $k=\lfloor n/2 \rfloor$) low-order interactions to construct a candidate pool (with at most $n^2/2$ interactions) first and the high-order interactions are chosen from this pool. This method is applied in both AutoFIS and AIM.

In the first two stages, the selected architecture parameters are optimized by GRDA optimizer and the other parameters are optimized by Adam optimizer. In the re-train stage, all the parameters are optimized by Adam optimizer.

Inspired by the transferability mentioned before, we sequentially introduce an MLP layer to learn more information and promote the performance in the re-train stage of AIM while this MLP layer does not exist in the search stage.

\begin{table}[t]
	\vspace{-0.6em}
	\caption{ Performance with different maximum interaction orders. \\``FI ratio'' is the remained ratio of feature interaction (FI).}
	\vspace{-1em}
	\label{tab:top}
	\centering
	\resizebox{0.38\textwidth}{!}{
		\begin{tabular}{c|ccc|ccc}
			\toprule
			\multicolumn{1}{c|}{\multirow{2}{*}{Model}}& \multicolumn{3}{c|}{Avazu} & \multicolumn{3}{c}{Criteo}\\
			& AUC & log loss& FI ratio & AUC & log loss & FI ratio \\ \midrule
			AutoFIS($2^{nd}$)   & 0.7852 & 0.3765 & 18\%  & 0.8007 & 0.5406 & 9\%  \\ 
			{AIM($2^{nd}$)}  & 0.7891  & 0.3742 & 16\% & 0.8022  & 0.5388 &  22\% \\  \midrule
			{AutoFIS($3^{rd}$)}  & 0.7870 & 0.3755 & 5\% & 0.8010  & 0.5404 & 2\%  \\ 
			{AIM($3^{rd}$)}   & 0.7912 & 0.3730  & 2\%  &  0.8029 & 0.5381 & 1\% \\  \midrule
			{AutoFIS($4^{th}$)}   & 0.7883 & 0.3748 & 0.6\% &  0.8012 & 0.5402 & 0.03\%   \\ 
			{AIM($4^{th}$)}  & 0.7920  & 0.3727  & 0.3\% & 0.8030 & 0.5379 & 0.09\%   \\  
			
			\bottomrule
		\end{tabular}
	}
\vspace{-0.6em}
\end{table}

\subsection{Overall Performance (RQ1)}

Table~\ref{tab:performance_overall_final} shows the overall performance of AIM and baselines on three datasets. Table~\ref{tab:top} summarizes the performance and FI ratio (i.e., percentage of selected feature interaction) with different maximum interaction order in AutoFIS and AIM.
From these two tables, we have the following observations. 
\begin{enumerate}[leftmargin = 12 pt]
    \item Compared with the \textbf{Factorization Models} and \textbf{Other Interaction Models}, AIM has achieved significant improvement. Among these base models, IPNN leverages an MLP to model high-order implicit information over the feature embeddings and interaction, mostly achieving better performance than the others. However, AIM further improves IPNN in terms of AUC by 0.66\% in Avazu and 0.21\% in Criteo via performing automatic search of IFs and embedding dimensions.
	\item In comparison with the existing representative \textbf{AutoML}-based models, AIM achieves the best performance. AutoFIS and AutoGroup only consider selecting useful feature interactions, ignoring the effectiveness of different IFs. Besides, none of these AutoML-based models take the embedding dimension search into consideration. In contrast, AIM proposes three components (i.e., FIS, IFS and EDS) to select significant feature interactions, appropriate IFs and necessary embedding dimensions automatically in a unified framework.  
	\item From Table ~\ref{tab:top}, we can observe that about 80\% of the $2^{nd}$-order interactions can be removed. As for high-order feature interaction selection, only about 2\% of all the $3^{rd}$-order feature interactions and about 0.3\%  of all the $4^{th}$-order feature interactions are selected with significant performance improvement. With a small amount of high-order interactions integrated, the relative performance improvement of AIM ($3^{rd}$-order) over AIM ($2^{nd}$-order) is 0.27\% and that of AIM ($4^{th}$-order) over AIM ($2^{nd}$-order) is 0.37\% in terms of AUC.
	
\end{enumerate}

\subsection{Ablation Study (RQ2)}\label{sec:exp_ablation_study}

In this subsection, we will first present the effectiveness of different components (namely, FIS, IFS and EDS) in AIM. 
Then we conduct experiments on real data to analyze why these three components are valid.

\subsubsection{Effectiveness of different components in AIM}

To validate the effectiveness of individual components in AIM, we propose several variants. Recall that AIM has three core components: FIS, IFS, EDS, and we apply experiments on these variants to verify the effectiveness of these components. The relationship among different variants and their performance is presented in Table~\ref{tab:var_comparison}, from which we can get the following conclusions.

\begin{table}[h]
	\vspace{-0.6em}
	\caption{ Performance comparison of different variants.}
	\vspace{-1em}
	\label{tab:var_comparison}
	\centering
	\resizebox{0.4\textwidth}{!}{
		\begin{threeparttable}
			{
		\begin{tabular}{c|ccc|cc|cc}
			\toprule
			\multicolumn{1}{c|}{\multirow{2}{*}{Model}}&
			\multicolumn{3}{c|}{Component} &\multicolumn{2}{c|}{Avazu} & \multicolumn{2}{c}{Criteo}\\
			  & FIS&IFS&EDS&AUC & log loss   & AUC & log loss  \\ \midrule
			 DeepFM & - & - & -&0.7836  & 0.3776 &  0.7991  & 0.5423 \\
			 AIM-IFS-EDS& $\surd$ & $\times$ & $\times$ & 0.7869 &  0.3755 &0.8015 & 0.5399  \\
			 AIM-EDS& $\surd$ & $\surd$ & $\times$ & 0.7894 & 0.3740  & 0.8024 & 0.5387  \\
			 AIM& $\surd$ & $\surd$ & $\surd$ & 0.7891 & 0.3742  &  0.8022 &0.5388  \\
			\bottomrule
		\end{tabular}
}
\end{threeparttable}
	}
	\vspace{-0.6em}
\end{table}

\begin{enumerate}[leftmargin = 12 pt]
 \item Comparing AIM-IFS-EDS with DeepFM, we can observe that removing those useless interactions can not only simplified the model but also significantly boost the prediction accuracy. The relative performance improvement of AIM-IFS-EDS over DeepFM is 0.42\% and 0.30\% for Avazu and Criteo dataset respectively in terms of AUC, which demonstrates the effectiveness of FIS component. 
 \item Besides, the performance gap between AIM-EDS and AIM-IFS-EDS indicates that selecting proper IFs for each feature interaction conduces to better performance, which is the functionality of IFS component.
 \item Comparing AIM-EDS and AIM, we can find that introducing EDS component in AIM for pruning redundant dimensions will not sacrifice the performance distinctly. However, EDS reduces model parameters significantly, by $\times$2.79 and $\times$3.16 in Avazu and Criteo dataset respectively, which will be shown in Section~\ref{sec:effective_eds}.
 
\end{enumerate}

\subsubsection{The Effectiveness of Selected Feature Interactions by FIS component}

To see how well $\boldsymbol{\alpha}$ values are learned in FIS component, we analyze the relationship between $\boldsymbol{\alpha}$ values and \textbf{$\mbox{statistics}\_\mbox{AUC}$}.

We define \textbf{$\mbox{statistics}\_\mbox{AUC}$} to represent the importance of a feature interaction to the final prediction. For a given interaction, we construct a predictor only considering this interaction where the prediction of a test instance is the statistical CTR ($\#\texttt{downloads} / \#\texttt{impressions}$) of specified feature interaction in the training set. Then the AUC of this predictor is defined as $\mbox{statistics}\_\mbox{AUC}$ with respect to this given feature interaction. Higher $\mbox{statistics}\_\mbox{AUC}$ indicates a more important role of this feature interaction in prediction.

\begin{figure}[h]
	\centering
	\vspace{-1.1em}
	\includegraphics[width=0.35\textwidth]{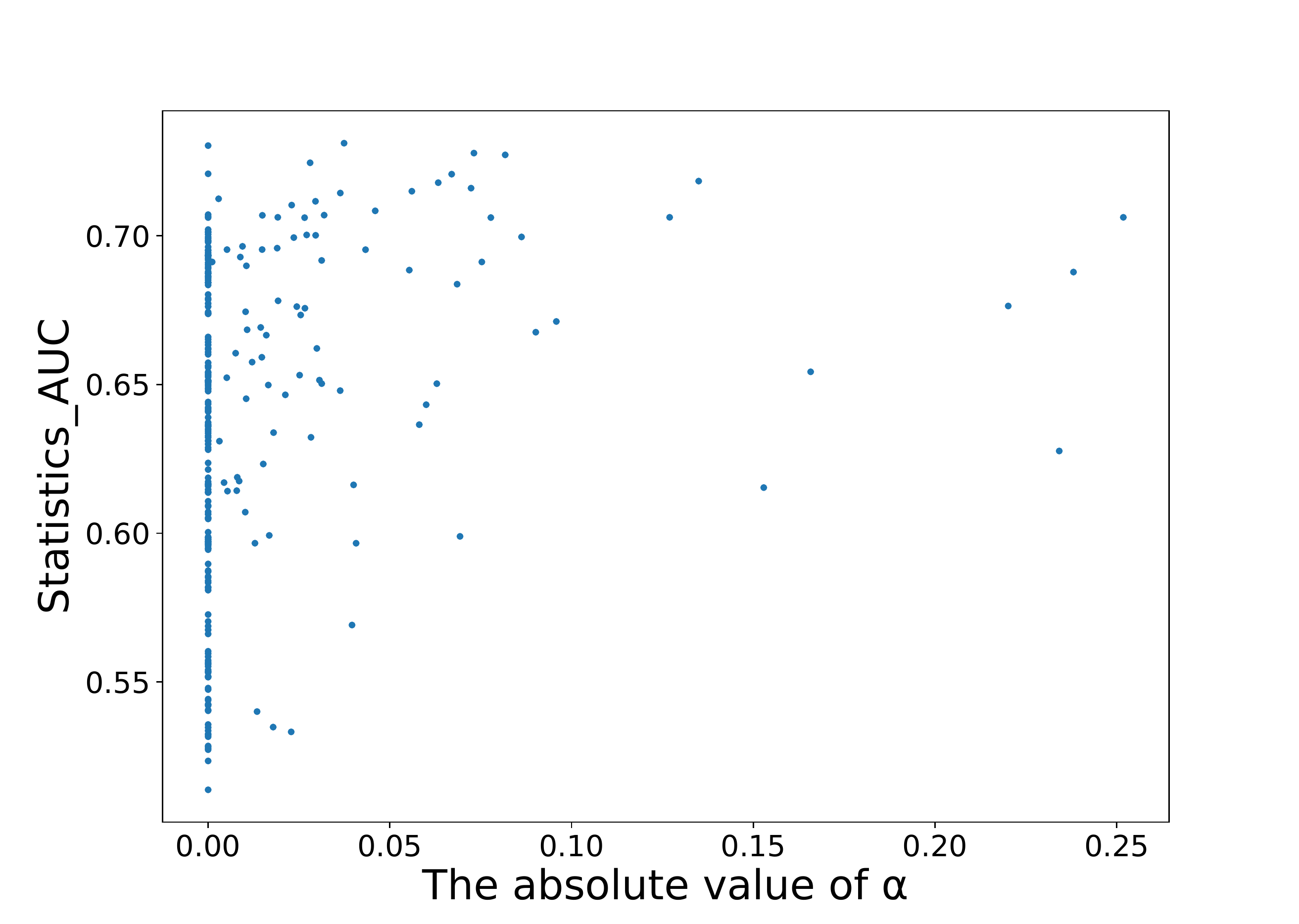}
	\vspace{-1em}
	\caption{ Relationship between $\mbox{statistics}\_\mbox{AUC}$ and $\alpha$ value for each $2^{nd}$-order interaction}
	\label{fig:auc_alpha}
	\vspace{-0.6em}
\end{figure}

As shown in Figure~\ref{fig:auc_alpha}, we can find that most of the feature interactions selected by FIS component (with high absolute $\boldsymbol{\alpha}$ value) have high $\mbox{statistics}\_\mbox{AUC}$, but not all feature interactions with high $\mbox{statistics}\_\mbox{AUC}$ are selected. That is because the information in these interactions may already exist in other interactions which are selected by our model.


To evaluate the effectiveness of the selected interactions by FIS component, we select the top $N$ $2^{nd}$-order feature interactions by FIS component and by $\mbox{statistics}\_\mbox{AUC}$ separately. We re-train two models with these two sets of interactions and compare their performance. The experimental result shows that, compared with the model by $\mbox{statistics}\_\mbox{AUC}$ with the same computational cost, the model with selected interactions by FIS component has improved AUC from 0.7804 to 0.7831 (log loss from 0.3794 to 0.3778), which demonstrates the superiority of FIS component.

\subsubsection{The Effectiveness of Selected Interaction Functions by IFS component}

\begin{figure}
\vspace{-0.5em}
	\centering
	\subfigure{
		\label{fig:r1} 
		\includegraphics[width=0.24\textwidth]{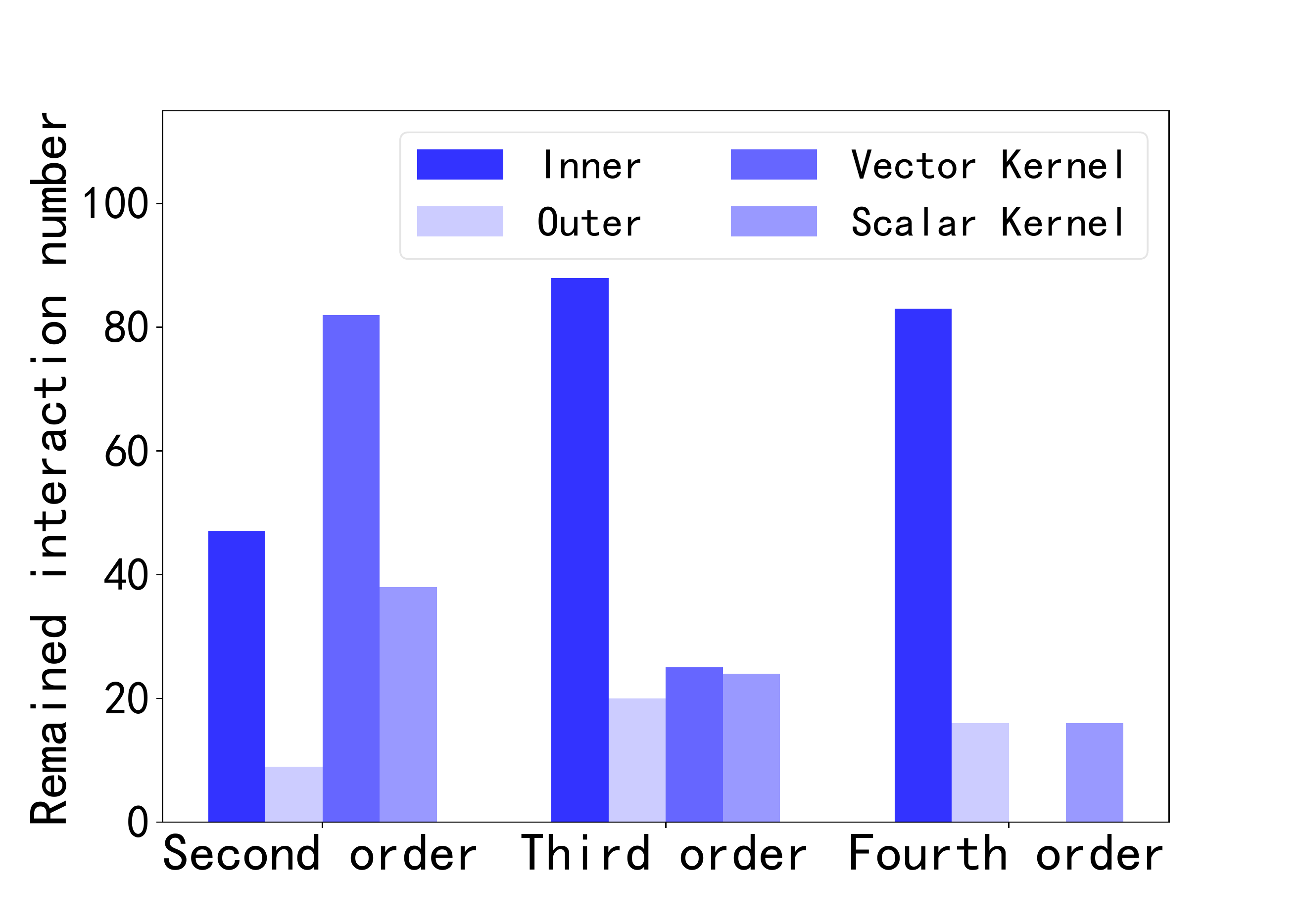}}
 	\hspace{-1em}
	\subfigure{
		\label{fig:r2} 
		\includegraphics[width=0.24\textwidth]{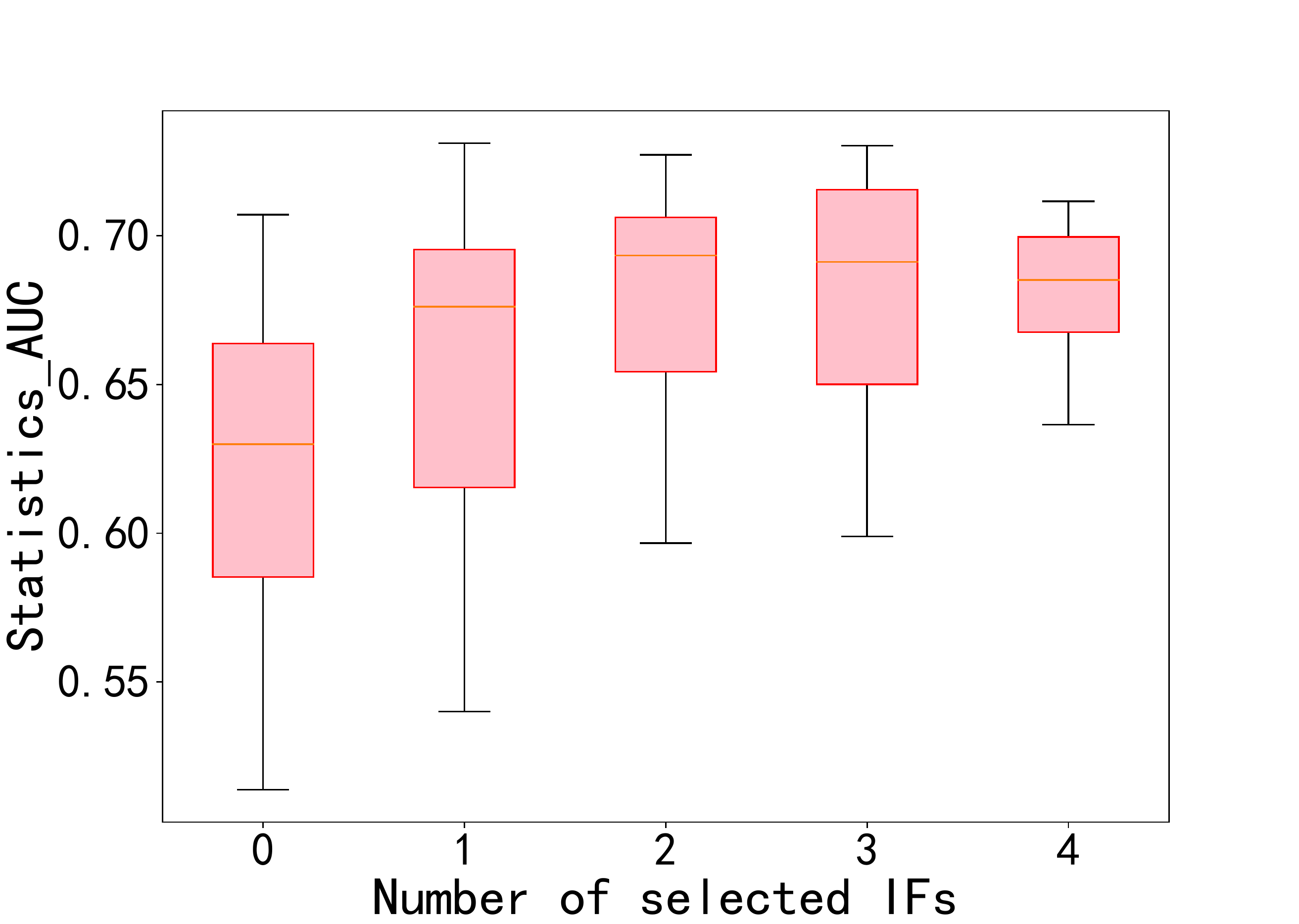}}
	\caption{The Effectiveness of Selected Interaction Functions by IFS. The left sub-graph is the selection frequency of each IF while the right sub-graph is the relationship between $\mbox{statistics}\_\mbox{AUC}$ and number of selected IFs for each $2^{nd}$-order interaction.}
	\vspace{-0.6em}
	\label{fig:IFS_effectiveness}
\end{figure}

To demonstrate that IFS component is able to find suitable IFs for individual interactions, we depict the number of selected IFs with respect to different orders of interactions on Avazu dataset, presented in the left sub-graph of Figure~\ref{fig:IFS_effectiveness}. The following observations can be concluded. 
(1) Each IF occupies a certain proportion, which verifies the need of different IFs for individual interactions. The performance superiority of IFS component validates that suitable IFs are identified for different interactions.  (2) Inner product takes the majority for high-order interactions and ranks at the second place for $2^{nd}$-order interaction. (3) Vector kernel product is the most popular IF for $2^{nd}$-order interaction but is unpopular for high-order interaction. (4) Scalar kernel product and outer product take only a small part of the overall selection, but still hold an indispensable part. 

Furthermore, we explore the relationship between the number of selected IFs and $\mbox{statistics}\_\mbox{AUC}$ for each feature interaction, which is presented in the right sub-graph of Figure~\ref{fig:IFS_effectiveness}. The figure indicates that more important feature interactions are likely to retain more IFs. Moreover, as the number of selected IFs increases, the improvement of $\mbox{statistics}\_\mbox{AUC}$ is gradually disappeared. This is reasonable, because it is more likely to be overfitting when more IFs are performed for a feature interaction. In other words, it is not the best to keep all the IFs. Instead, selecting suitable IFs for individual interactions is more appropriate.



\subsubsection{The Effectiveness of Selected Embedding Dimensions by EDS component}\label{sec:effective_eds}

In this section, we analyze why the reserved embedding size of a given feature determined by EDS component is reasonable. Before that, we first show the effectiveness of EDS.

As mentioned in Section~\ref{sec:autoembgate}, function-wise embeddings (FWE) improve the performance compared with shared embedding (SE), therefore we compare the performance of AIM-EDS (SE) and AIM-EDS (FWE). 
As the model size is significantly larger if the embedding size of FWE keeps the same as that of SE, which is the motivation of proposing EDS component. We further compare AIM (FWE) and AIM-EDS (FWE) to validate the effectiveness of EDS. 

\begin{table}[h]
	\vspace{-0.6em}
	\caption{ Performance and parameters comparison. ``params'' does not include the pruned parameters and it is count as $10^7$.}
	\vspace{-1em}
	\label{tab:owe_se_all}
	\centering
	\resizebox{0.42\textwidth}{!}{
		\begin{threeparttable}
			{
		\begin{tabular}{c|ccc|ccc}
			\toprule
			\multicolumn{1}{c|}{\multirow{2}{*}{Model}}& \multicolumn{3}{c|}{Avazu} & \multicolumn{3}{c}{Criteo}\\
			  & AUC & log loss &  params  & AUC & log loss &  params \\ \midrule
			AIM-EDS (SE) & 0.7870 & 0.3754 & 2.7& 0.8015 & 0.5399 & 2.8\\
			AIM-EDS (FWE) & 0.7894 & 0.3740  &  10.6  & 0.8024  & 0.5387 & 9.8\\
			AIM (FWE) & 0.7891 & 0.3742 &3.8& 0.8022 & 0.5388 & 3.1\\
			\bottomrule
		\end{tabular}
}
\end{threeparttable}
	}
	\vspace{-0.6em}
\end{table}
The above mentioned experimental comparisons are presented in Table~\ref{tab:owe_se_all}, from which We can get the following conclusions.
\begin{enumerate}[leftmargin = 12 pt]
	\item The comparison between AIM-EDS (SE) and AIM-EDS (FWE) verifies that multiple embeddings for different IFs can boost the performance but lead to much more parameters.
	\item Comparing AIM (FWE) with AIM-EDS (FWE), equipping EDS component achieves almost the same high performance as AIM-EDS (FWE) but needs only one third of the parameters required by AIM-EDS (FWE). The parameters of AIM-EDS (FWE) are slightly more than that of AIM-EDS (SE) in a comparable amount.
\end{enumerate}


To see the reserved embedding size of each field, we summarize these numbers in Table~\ref{tab:embedding_size}, together with the vocabulary size of each field (i.e., the number of features in a field) on Avazu dataset.
The correlation between vocabulary size and reserved embedding size presented in Table~\ref{tab:embedding_size} indicates that the fields with larger vocabulary size tend to achieve larger embedding size. That is because these fields may be more informative and predictive to the task. However, as can also be observed, vocabulary size is not the only factor to determine the embedding size. Other unknown factors need to be learned by machine learning models, such as EDS component in AIM.

\begin{table}[h]
	\centering
		\vspace{-0.6em}
	\caption{ Correlation between vocabulary size and reserved embedding size}
		\vspace{-1em}
		\resizebox{0.42\textwidth}{!}{
\begin{tabular}{cc|cc|cc}
	\toprule
	\tabincell{c}{vocabulary} &  embedding & vocabulary & embedding & vocabulary  & embedding \\\midrule
	   4 & 1  &  10  & 25  &  426  & 54 \\
	   4 & 11  & 24   &  6 &  2417 & 33 \\
	   5 & 1  &  24  &  35 &  3135 & 55 \\
	   7 & 1  &  28  &  5 &   3487 & 56 \\
	   7 & 8  &  60  &  23 &  4002 & 55 \\
	   7 & 16  & 67   & 6  &  5925 & 81 \\
	   8 & 1  &  166  &  32 & 101449  & 22 \\
	   9 & 1  &  252  &  15 & 523672  & 98 \\
	\bottomrule
\end{tabular}
}
	\vspace{-0.6em}
	\label{tab:embedding_size}
\end{table}

\subsection{Space and Time Complexity (RQ3)}\label{sec:exp_space_time}

In this section, we discuss the space and time complexity of AIM with $2^{nd}$-order interactions and high-order interactions, respectively.
In order to facilitate calculation, we choose Avazu (with 645,195 features) to analyze the memory usage, training and inference efficiency of AIM in comparison with other models. To compare the training and inference efficiency, we train these models for the whole training set in one epoch and test these models with 200,000 instances (batch size=2,000) on an NVIDIA 1080Ti GPU.

\subsubsection{Complexity Analysis for $2^{nd}$-order interaction}

First, we analyze the complexity of all the models with $2^{nd}$-order interaction modeling and present the results in Table~\ref{tab:para_time}. 
Note that we only consider the parameters and training time in the re-train stage of AutoFIS and AIM model.
Because the useful feature interactions, effective interaction-IF pairs and reserved embedding dimensions can be searched once for all, while the model training over new data only needs to perform the re-train stage based on the searched results in the first two search stages.
To complete the complexity analysis of AIM, the complexity analysis of the search stages is presented in Section~\ref{sec:complexity-high-order}.

\begin{table}[h]
	\vspace{-0.5em}
	\caption{ Space and time complexity comparison of models with $2^{nd}$-order interaction. ``train'' is the training time for the whole training set in one epoch and ``inference'' is the inference time for 200,000 samples.}
	\vspace{-0.5em}
	\label{tab:para_time}
	\centering
	\resizebox{0.32\textwidth}{!}{
	\begin{threeparttable}[h]{
		\begin{tabular}{c|ccc}
			\toprule
			Model &  params ($10^7$) &train (min)& inference (s)  \\ \midrule
			FM & 2.6           &3.3 & 0.53    \\
			FFM  & 6.0         &3.7  & 0.25  \\
			DeepFM & 2.9        &5.0 & 0.88   \\
			IPNN & 2.9         &5.5 & 0.96   \\
			AutoFIS & 2.9   &4.1 & 0.59   \\
			AIM-EDS & 10.6     &8.7 & 0.81   \\
			AIM &3.8  &9.0  & 0.83   \\
			\bottomrule
		\end{tabular}
		}
		\end{threeparttable}
	}
	\vspace{-0.2em}
\end{table}

As for the space complexity, the number of parameters of AutoFIS and its base model (DeepFM) are very similar. AIM-EDS consumes the largest amount of memory as it utilizes multiple embeddings for each feature. By integrating EDS component, AIM reduces the amount of parameters to a comparable level as other models without sacrificing model performance (as already discussed in Section~\ref{sec:effective_eds}).

In terms of training and inference efficiency, we find that AutoFIS is apparently faster than the base model (DeepFM) both in training and inference time, because removing those useless interactions contributes to accelerating the model training and inference, with higher performance. Although the training time of AIM is acceptably larger than the other models, the inference of AIM is very efficient (i.e., even more efficient than DeepFM). Such time complexity makes it possible to deploy AIM in an industrial system.

\subsubsection{Complexity Analysis for high-order interaction}\label{sec:complexity-high-order}

In this section, we analyze the complexity of AIM with high-order interactions. We consider the training and inference time as time complexity. The training time contains three parts: \emph{search interaction-IF stage}, \emph{search embed stage} and \emph{re-train stage}. Also, we take the parameters into account to compare space complexity.

\begin{table}[h]
	\vspace{-0.8em}
	\caption{ Space and time complexity for AIM with high-order interactions. ``train'' is the training time for the whole training set in one epoch and ``inference'' is the inference time for 200,000 samples.}
	\vspace{-0.7em}
	\label{tab:para_high_order}
	\centering
	\resizebox{0.35\textwidth}{!}{
	\begin{threeparttable}{
		\begin{tabular}{c|ccc}
			\toprule
			Model & params ($10^7$) & train (min) & inference(s) \\ \midrule
			AIM($2^{nd}$) & 3.8 & 9.0 &0.83\\
			AIM($3^{rd}$) & 7.1 &14.8 &1.35\\
			AIM($4^{th}$) & 9.9  &19.5 &1.72\\
			\bottomrule
		\end{tabular}
		}
		\end{threeparttable}
	}
	\vspace{-0.2em}
\end{table}

\begin{table}[h]
	\vspace{-0.2em}
	\caption{ Training time of AIM for different stages. It is the training time for the whole training set in one epoch and counted by min.}
	\vspace{-0.7em}
	\label{tab:time_diff}
	\centering
	\resizebox{0.35\textwidth}{!}{
	\begin{threeparttable}{
		\begin{tabular}{c|ccc}
			\toprule
			Model & search interaction-IF &  search embed & re-train \\ \midrule
			AIM($2^{nd}$) & 13.0 & 9.2 & 9.0\\
			AIM($3^{rd}$) & 20.5 & 17.1 & 14.8 \\
			AIM($4^{th}$) & 29.4 & 22.2  & 19.5 \\
			\bottomrule
		\end{tabular}
		}
		\end{threeparttable}
	}
	\vspace{-1.2em}
\end{table}


From Table~\ref{tab:para_high_order} and Table~\ref{tab:time_diff} we can find that, thanks to the selection algorithm for high-order feature interaction proposed in Section~\ref{sec:autofis} which reduces the time complexity of exploring $p^{th}$-order interactions from $O(n^p)$ to $O(n^2)$, the overhead of introducing high-order interactions in terms of training time and space complexity is acceptable. 
It takes about 155 minutes in total for AIM to search important $2^{nd}$-, $3^{rd}$- and $4^{th}$-order feature interactions with appropriate IFs and proper embedding sizes for each field with \emph{a single GPU}.
The same decisions will take the human engineers dozens of days or weeks to make by analysis and experiments.

\subsection{Transferability of the Selected Interaction-IF pairs by AIM (RQ4)}\label{sec:exp_transfer}

\begin{table}[h]
	\centering
	\vspace{-0.9em}
	\caption{ Performance of transferring interaction-IF pairs selected by AIM to other models.}     
	\vspace{-0.5em}
	\label{tab:performance_ipnn}
	\centering
	\resizebox{0.33\textwidth}{!}{
		\begin{tabular}{c|cc|cc}
			\toprule
			\multicolumn{1}{c|}{\multirow{2}{*}{Model}}& \multicolumn{2}{c|}{Avazu} & \multicolumn{2}{c}{Criteo}\\
			\multicolumn{1}{c|}{} & AUC & log loss   & AUC & log loss \\
			\midrule
			\multicolumn{1}{c|}{DeepFM}  &  0.7836 & 0.3776  & 0.7991 & 0.5423     \\ 
			\multicolumn{1}{c|}{DeepFM+IF($2^{nd}$)}  & 0.7858  & 0.3762  & 0.7999 & 0.5415     \\
			\multicolumn{1}{c|}{DeepFM+IF($3^{rd}$)}  & 0.7868  & 0.3758  & 0.8004 & 0.5411     \\
			\multicolumn{1}{c|}{DeepFM+IF($4^{th}$)}  & \textbf{0.7872}  &  \textbf{0.3755} & \textbf{0.8005} & \textbf{0.5410}     \\
			\midrule
			\multicolumn{1}{c|}{PNN}   &  0.7868 & 0.3756  & 0.8013 & 0.5401        \\ 
			\multicolumn{1}{c|}{PNN+IF($2^{nd}$)}  &  0.7893 & 0.3741  & 0.8024 & 0.5387     \\
			\multicolumn{1}{c|}{PNN+IF($3^{rd}$)}  &  0.7915 & 0.3728  & 0.8028 & 0.5382      \\
			\multicolumn{1}{c|}{PNN+IF($4^{th}$)}  &  \textbf{0.7921} & \textbf{0.3725}  & \textbf{0.8029} & \textbf{0.5381}     \\
			\bottomrule
		\end{tabular}
	}
	\vspace{-0.7em}
\end{table}

In this subsection, we investigate whether the important feature interactions with appropriate IFs learned by AIM could be transferred to other models for boosting their performance. 
We apply the searched interaction-IF pairs to two benchmark methods, i.e., DeepFM~\cite{deepfm} and PNN~\cite{pnn}, to explore its transferability. 

As shown in  Table~\ref{tab:performance_ipnn}, using $2^{nd}$-order feature interactions and IFs selected by our model achieves much better performance in both DeepFM and IPNN, with around 16\% of all the $2^{nd}$-order interactions in Avazu dataset and around 20\% in Criteo dataset. Moreover, the promotion is more significant by using high-order interactions with IFs. More specifically, with $4^{th}$-order interactions and corresponding IFs, the performance is improved by 0.46\% for DeepFM and 0.67\% for PNN, respectively. 
Both evidences verify the transferability of the selected feature interactions with chosen IFs in AIM.

\subsection{Deployment \& Online Experiments (RQ5)}

Online experiments are conducted in the recommender system of a mainstream app market to verify the superior performance of our model AIM, where hundreds of millions of daily active users generate hundreds of billions of user log events every day in the form of implicit feedback such as browsing, clicking and downloading apps.
In the online serving system, hundreds of candidate apps that are most likely to be downloaded by the users are retrieved from the universal app pool. Then these candidate apps are ranked by a fine-tuned ranking model (such as DeepFM, AIM) before presenting to users. To guarantee user experience, the overall latency of the above-mentioned candidate selection and ranking is required to be within a few milliseconds.
Because of the longer training time and more complicated functions of AIM, we simplify it and put a simplified version to deploy on our recommender system for the trade-off between efficiency and performance. 
The commercial model is deployed in a cluster, where each node is with 48 core Intel Xeon CPU E5-2670 (2.30GHZ), 400GB RAM and as well as 2 NVIDIA TESLA V100 GPU cards.

Specifically, a three-week A/B test is conducted in a major list of an app recommendation scenario. Our baseline in online experiments is DeepFM, which is a strong baseline due to its extraordinary accuracy and high efficiency. 
For the control group, users are randomly selected and presented with recommendation results generated by DeepFM. 
For the experimental group (7\% users), users are presented with recommendation apps generated by our AIM model.

\begin{figure}[h]
	\centering
	\vspace{-1.1em}
	\includegraphics[width=0.45\textwidth]{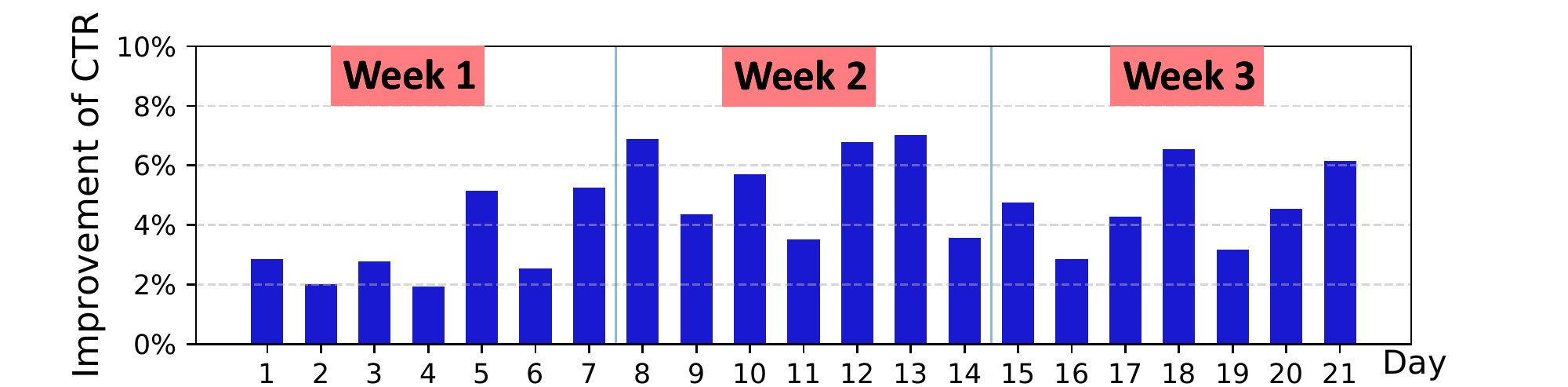}
	\vspace{-1.1em}
	\caption{ Online experimental results of CTR.}
	\vspace{-0.5em}
	\label{fig:online_ctr}
\end{figure}

Figure~\ref{fig:online_ctr} shows the improvement of the experimental group over the control group in terms of CTR ($\#\texttt{downloads} / \#\texttt{impressions}$). 
We can observe that the average improvement of CTR is \textbf{4.4\%} (statistically significant), which brings enormous commercial profits. These results demonstrate the magnificent effectiveness of our proposed model in industrial applications.

%% file: conclusion.tex
\section{Conclusion}

In this work, we propose Automatic Interaction Machine (AIM) with three core components, namely, Feature Interaction Search (FIS), Interaction Function Search (IFS) and Embedding Dimension Search (EDS) to automatically select significant feature interactions, appropriate IFs and necessary embedding dimensions in a unified framework. The proposed AIM is easy to implement with marginal search costs, and the performance improvement is significant in two benchmark datasets and one private dataset. Our model has been deployed onto the training platform of a mainstream app market recommendation service, with significant economic profit demonstrated. 

\vspace{-1.0em}